\documentclass[aps,prd,amsmath,amssymb,superscriptaddress, preprintnumbers,preprint,nofootinbib,a4paper]{revtex4}
\pdfoutput=1

\usepackage[utf8x]{inputenc}
\usepackage{amsmath}
\usepackage{amsfonts}
\usepackage{amssymb}
\usepackage{graphicx,amsfonts,color,comment,amsmath,hyperref,float}
\usepackage{bm}% bold math
\usepackage{cancel}
\usepackage{enumitem}

\usepackage[normalem]{ulem}

\usepackage{color}

\newcommand{\beq}{\begin{eqnarray}}
\newcommand{\eeq}{\end{eqnarray}}
\newcommand{\GeV}{\,\text{GeV}}
\newcommand{\TeV}{\,\text{TeV}}

\newcommand{\ifb}{\,{\rm fb}^{-1}}

\newcommand{\ttb}{t\bar{t}}

\def\Sherpa{\textsc{Sherpa}}
\def\Comix{\textsc{Comix}}
\def\Recola{\textsc{Recola}}
\def\Collier{\textsc{Collier}}

\def\mg5{\textsc{MadGraph\_aMC\@NLO}}
\def\Pythia8{\textsc{Pythia8}}
\def\Rivet{\textsc{Rivet}}
\def\Feynrules{\textsc{FeynRules}}

\def\eq#1{{eq.~(\ref{#1})}}

\def\fig#1{{fig.~(\ref{#1})}}
\def\sec#1{{sec.~(\ref{#1})}}

\def\app#1{{app.~(\ref{#1})}}

\begin{document}

%\title{Probing high mass resonances in boosted top-quark pair production}
%\title{Interpreting data as resonances decaying into top-quark pairs}
\title{Constraining scalar resonances with top-quark pair production at the LHC}

\author{Diogo Buarque Franzosi}
\affiliation{
 Institut f\"ur Theoretische Physik, Universit\"at G\"ottingen, Friedrich-Hund-Platz 1, 37077 G\"ottingen, Germany}

\author{Federica Fabbri}
\affiliation{Università and INFN Bologna} 

\author{Steffen Schumann}
\affiliation{
 Institut f\"ur Theoretische Physik, Universit\"at G\"ottingen, Friedrich-Hund-Platz 1, 37077 G\"ottingen, Germany}
 
\preprint{MCnet-17-17}

\begin{abstract}

  Constraints on models which predict resonant top-quark pair production at the LHC are provided via
  a reinterpretation of the Standard Model (SM) particle level measurement of the top-anti-top invariant
  mass distribution, $m(\ttb)$. We make use of state-of-the-art Monte Carlo event simulation to perform
  a direct comparison with measurements of $m(\ttb)$ in the semi-leptonic channels, considering both the
  boosted and the resolved regime of the hadronic top decays. A simplified model to describe various
  scalar resonances decaying into top-quarks is considered, including CP-even and CP-odd, color-singlet
  and color-octet states, and the excluded regions in the respective parameter spaces are provided.

\end{abstract}
 
\maketitle

%%%%%%%%%%%%%%%%%%%%%%%

\tableofcontents

\section{Introduction}

With its mass being close to the electroweak scale the top quark is very special. It might intimately
be connected to the underlying mechanism of electroweak symmetry breaking (EWSB). Consequently, studying
top-quark production and decays at colliders might provide a portal to New Physics (NP). The
Large Hadron Collider (LHC), providing proton--proton collisions currently at 13 TeV centre-of-mass energy,
can be seen as a top-quark factory. It allows to search for anomalous top-quark production and decay processes,
considered as low energy modifications of the Standard Model (SM) parametrized by effective operators~\cite{Barger:2011pu,Choi:2012fc,Franzosi:2015osa,Zhang:2016omx,Englert:2016aei,Cirigliano:2016nyn}, or,
as the direct production of intermediate resonances, which have been hunted for a long time at different
experiments~\cite{Aaltonen:2009tx,Khachatryan:2015sma,Chatrchyan:2013lca}. 

Heavy scalar resonances that decay into a pair of top quarks are predicted by several NP scenarios,
in particular the Two Higgs Doublet Model (THDM), supersymmetric theories and models of dynamical EWSB.
In this paper, we provide a framework to reinterpret the SM $\ttb$ differential cross section measurements as exclusion limits for signatures of NP resonances decaying into $\ttb$.
The framework relies on the comparison between particle-level data with state-of-the-art event simulation
and the interpretation of deviations in terms of NP models.  It is based on four main ingredients
\begin{enumerate}
\item 
A Monte Carlo event generator which allows the precise and realistic description of particle-level observables. 
 
In order to theoretically describe top-quark pair production at the LHC, we make use of state-of-the-art
event simulations provided by the \Sherpa~\cite{Gleisberg:2008ta} event-generator framework. This implies
the usage of techniques to match leading and next-to-leading order QCD matrix elements with parton showers 
and merging different parton-multiplicity final states.
\item The precise measurement of SM processes from fiducial kinematical regions provided as differential
  particle-level observables by LHC experiments, and available through the \textsc{Rivet}
  package~\cite{Buckley:2010ar}. Here we used the ATLAS analyses of top-quark pair production in the
  boosted~\cite{Aad:2015hna} and resolved~\cite{Aad:2015mbv} regimes.
\item A general parametrization of NP whose predictions for colliders can be computed efficiently.
  We adopt a Lagrangian which describes scalar resonances that can be CP-even or odd and color singlet
  or octet. We devise a \emph{reweighting} method to describe the model prediction  in the $m(\ttb)$
  distribution for a wide range of the parameter space in a fast and efficient manner. 
\item A statistical interpretation to decide what regions of parameter space of the model are ruled
  out at a given confidence level. We adopt here a simplified $\chi^2$ analysis.
\end{enumerate}
A similar method to constrain NP with SM measurements in several other channels has recently been
presented in Ref.~\cite{Butterworth:2016sqg}.
  These approaches are complementary to model-specific searches in the respective final states. 
  They provide systematic methods for the theory community to derive more realistic exclusion limits
  for any particular model, not relying on the experiment-specific assumptions. 

In the rest of the paper we explain these 4 points in detail.
In Sec. II we describe the set-up of our event simulation. In Sec. III
we give details on the analyses used in the boosted and the resolved regime and validate our SM predictions
by comparing them to experimental data. In Sec. IV we introduce our simplified model of beyond the SM scalar
resonances and describe the implementation in our simulation framework, based on an event-by-event \emph{reweighting}.
In Sec. V we present a statistical analysis to assess the region in parameter space accessible by the LHC 
experiments and provide interpretations in terms of some specific models. We finally conclude in Sec. VI.

\section{Simulation framework}
\label{sec:simulation}

When searching for imprints of resonant contributions in top-quark pair production at the LHC, a detailed
understanding of the SM production process is vital. In particular, as there are non-trivial interference
effects between NP signals and SM amplitudes that determine the shape of the resulting top-pair invariant-mass
distribution. In order to obtain realistic and reliable predictions for the top-pair production process,
we make use of state-of-the-art particle-level simulations, based on higher-order matrix elements matched
to parton-shower simulations and hadronization. 

Our analysis focuses on observables in the semi-leptonic decay channel of top-quark pair production, i.e.
\begin{equation}
pp\to \ttb \to b\bar{b}jj\ell\nu\text{ + jets}\,,
\label{eq:process}
\end{equation}
where $\ell$ denotes muons or electrons, $\nu$ the corresponding neutrinos, $b$ are bottom quarks and $j$
light quarks or gluons. These decay products and the associated radiation might be reconstructed as
well-separated objects, i.e. light-flavour jets, $b$-jets and a lepton, or, in the boosted regime, as a large-area jet,
containing the hadronic decay products, additional jets and a lepton. In either case, to realistically simulate
the associated QCD activity, higher-order QCD corrections need to be considered. 

To describe the SM top-pair production process we use the \Sherpa\ event-generation
framework~\cite{Gleisberg:2003xi,Gleisberg:2008ta}. We employ the techniques to match LO and NLO QCD matrix
elements to \Sherpa's dipole shower~\cite{Schumann:2007mg} and to merge processes of variable partonic
multiplicity~\cite{Hoeche:2009rj,Hoeche:2012yf}. Leading-order and real-emission correction matrix elements
are obtained from \Comix~\cite{Gleisberg:2008fv}. Virtual one-loop amplitudes, contributing at NLO QCD, are
obtained from the \Recola\ generator~\cite{Actis:2016mpe,Biedermann:2017yoi} that employs the \Collier\
library~\cite{Denner:2016kdg}. Top-quark decays are modelled at leading-order accuracy through \Sherpa's
decay handler, that implements Breit-Wigner smearing for the intermediate resonances and preserves spin
correlations between production and decay~\cite{Hoche:2014kca}. We treat bottom-quarks as
massive in the top-quark decays and the final-state parton-shower evolution~\cite{Krauss:2016orf}.

To validate the SM predictions we also consider leading-order simulations in the \mg5 framework~\cite{Alwall:2014hca}.
The hard-process' partonic configurations get showered and hadronized through \Pythia8~\cite{Sjostrand:2014zea}.
The spin-correlated decays of top quarks are implemented through the \textsc{MadSpin} package \cite{Artoisenet:2012st}.
Samples of different partonic multiplicity are merged according to the $k_T$-MLM prescription described in \cite{Alwall:2007fs}.

For the top-quark and $W$-boson, the following mass values are used
\begin{equation}
  m_t=172\;{\rm GeV}\,,\quad m_W=80.39\;{\rm GeV}\,,
\end{equation}
and the corresponding widths are calculated at leading order, assuming for the remaining electroweak input parameters
$m_Z=91.19\;{\rm GeV}$ and $G_\mu=1.16637\times 10^{-5}\;{\rm GeV}^{-2}$. In the following section
%, Sec.~\ref{sec:analysis},
we present a comparison of our simulated predictions against ATLAS measurements and discuss their systematics. Alongside, we give details on the QCD input parameters and calculational choices used there. 
%%%
%
%We used the PDF NNPDF30 LO\cite{} set with $\alpha_S(M_Z)=0.13$ for the LO computations and \com{???} for the NLO computation.
%For the 0 jet samples (without merging) we used the renormalization, $\mu_R$, and factorization, $\mu_F$, scale
%\begin{equation}
%\mu_R=\mu_F = 0.25 ( m_t^2 + 0.5 (p_{t}^2 + p_{\bar{t}}^2) )
%\end{equation}
%\com{For the merged samples we used the dynamical scales described in \cite{} and \cite{} for \Sherpa and \mg5 respectively.}

%\com{The different merging setups tried are shown in \sec{app:merging}.}

\section{Analysis framework}
\label{sec:analysis}

In what follows we describe the event selections used to identify the top-quark pair-production
process, used later on to study the imprint of resonant NP contributions. Thereby,
we closely follow the strategies used by the LHC experiments. Our simulated events from
\Sherpa\ and \mg5\ are produced in the \textsc{HepMC} output format \cite{Dobbs:2001ck} and
passed to \textsc{Rivet}~\cite{Buckley:2010ar} where we implement our particle-level selections. 

We consider two analyses, based on measurements performed using the ATLAS detector of the
differential $\ttb$ production cross sections in proton-proton collisions at $\sqrt{s} = 8 $ \TeV\
with an integrated luminosity of $L=20.3\ifb$ 
\cite{Aad:2015hna,Aad:2015mbv}.
Both analyses select events in the \emph{leptons+jets} decay channel. The two measurements indicated
in the following as \emph{Resolved} and \emph{Boosted} are optimized for different regions
of phase space. The \emph{Boosted} analysis, cf. Ref.~\cite{Aad:2015hna}, is designed to
enhance the selection and reconstruction efficiency of highly-boosted top quarks with transverse
momentum $p_T > $ 300 GeV, that might originate from the decay of a heavy resonance with
mass $m> 600\,{\rm GeV}$. In such events the decay products of the hadronic top overlap, due to
the high Lorentz boost. In turn, they cannot be reconstructed as three distinct jets. The
\emph{Resolved} analysis, based on Ref.~\cite{Aad:2015mbv}, measures the differential cross section
as a function of the full kinematic spectrum of the $\ttb$ system and is useful to identify and
reconstruct rather light resonances.

The selection requirements are applied on leptons and jets at particle level, i.e. after hadronization.
In our simulated data we discard any detector resolution, i.e. smearing effects. All the leptons used
in the analyses, i.e. $e, \mu, \nu_{e}$ and $\nu_{\mu}$ must not originate from hadrons, neither directly
nor through a $\tau$-lepton decay. In this way the leptons are guaranteed to originate from $W$-boson
decays without a specific matching requirement. The four-momenta of the charged leptons are modified by
adding the four-momenta of all photons found in a cone of $\Delta R$ = 0.1 around the leptons'
direction, thus representing dressed leptons. The missing transverse energy of the events ($E_T^{miss}$)
is defined from the four-vector sum of the neutrinos not resulting from hadron decays.

Jets are clustered using the anti-$k_{T}$ algorithm \cite{Cacciari:2008gp} with a radius of $R=0.4$ for
small-R jets and $R=1.0$ for the large-R jets, using all stable particles, excluding the selected
dressed leptons, as input. All small-R jets considered during the selections are required to have
$p_T > $ 25 GeV and $|\eta| < $ 2.5, while for large-R jets we demand $p_T > $ 300 GeV and $|\eta| < $ 2.
The small-R jets are considered $b$-tagged if a $b$-hadron with $p_T > $ 5 GeV is associated to the jet
through a ghost-matching procedure \cite{Cacciari:2008gn,Cacciari:2007fd}. To remove most of the
contribution coming from the interaction of the proton remnants, i.e. the underlying event, and to
reduce the dependence on the generator, large-R jets are groomed following a trimming procedure with
parameters $R_{sub} = 0.3$ and $f_{cut}$ = 0.05, for details of the procedure see Ref.~\cite{Krohn:2009th}.

Both the \emph{Resolved} and the \emph{Boosted} selections require a single lepton with $p_T > $ 25 GeV
and $|\eta| < $ 2.5. In the \emph{Resolved} analysis, apart from the leptons, the events are required to
have at least four small-R jets and at least two of them have to be $b$-tagged. In the \emph{Boosted}
analysis the events are required to have $E_T^{miss} > $ 20 GeV and $E_T^{miss}+m^W_{T}> 60\;{\rm GeV}$, with
$m^W_T = \sqrt{2 p_T^l E_T^{miss} (1 - \cos\Delta \phi)}$, the transverse mass of the leptonically decaying
$W$-boson, where $\Delta \phi$ denotes the azimuthal angle between the lepton and the $E_{T}^{miss}$ vector.
The presence of at least one small-R jet with $\Delta R ({\rm lepton},\,${\rm small-R jet})$<1.5$ is required.
In case more than one jet fulfills this requirement the jet with higher $p_T$ is considered as the
jet originating from the leptonic top decay, dubbed \emph{lep-jet} candidate. Furthermore, it is required the
presence of a trimmed large-R jet with mass $m_j^{R=1.0}> 100$ GeV and $\sqrt{d_{12} }> $ 40 GeV, where
$\sqrt{d_{12} }$ is the $k_t$ distance~\cite{Aad:2013gja,Butterworth:2002tt} between the two subjets in the
last step of the jet reclustering, \emph{i.e.} $\sqrt{d_{12}} = \min(p_{T1}, p_{T2})\,\Delta R_{1,2}$ . If more
than one large-R jet fulfills these requirements the one with highest transverse momentum is considered as
the \emph{had-jet} candidate. The \emph{had-jet} candidate must furthermore satisfy certain kinematic
requirements: $\Delta \phi({\emph{had-jet}},\,{\rm lepton}) > 2.3$ and
$\Delta R({\emph{had-jet}},\,{\emph{lep-jet}}) > 1.5$. The final requirement in the \emph{Boosted} selection
is that at least one $b$-tagged jet in $\Delta R ({\emph{had-jet}},\,{\rm jet}) < 1$ is found or
that the \emph{lep-jet} candidate is $b$-tagged. The \emph{Resolved} and \emph{Boosted} event selections
are summarized in Tab.~\ref{tab:cuts}.

\setlist{nolistsep}

\begin{table}[h!]
\centering
%\resizebox{\textwidth}{!}{
\begin{tabular}{p{0.33\textwidth}|p{0.33\textwidth}p{0.33\textwidth}}
\toprule
\multicolumn{3}{c} {event selections} \\
\hline
\multicolumn{3}{c} {Exactly one lepton ($\mu$ or $e$) with $p_T > 25$ GeV and $|\eta| < 2.5$ } \\
\hline
\emph{Resolved analysis}  & \multicolumn{2}{l}{\emph{Boosted analysis}} \\
\vphantom{nulla} & \multicolumn{2}{l}{$E_T^{miss} > 20$ GeV and $E_T^{miss}+m^W_{T} > $ 60 GeV} \\
$\ge$ 4 small-R jet:
\begin{itemize}[label={-}]
\item $p_T > 25$ GeV, $|\eta| < $ 2.5
\end{itemize}
& \multicolumn{2}{p{0.66\textwidth}}{$\ge$ 1 large-R jet:

\begin{itemize}[label={-}]
\item $p_T > 300$ GeV, $|\eta| < $ 2
\item $\sqrt{d_{12}} > $ 40 GeV
\item $m_j^{R=1.0} > 100$ GeV
\item $\Delta \phi($ large-R jet, lepton$) > 2.3 $
\end{itemize}}\\
& \multicolumn{2}{p{0.66\textwidth}}{$\ge$ 1 small-R jet:
\begin{itemize}[label={-}]
\item $p_T > 25$ GeV, $|\eta| < $ 2.5
\item $\Delta R ($lepton, small-R jet$) < 1.5$
\item $\Delta R($small-R jet, large-R jet$) > 1.5$
\end{itemize}}\\

$\ge$ 2 $b$-tagged jets & \multicolumn{2}{p{0.66\textwidth}}{$\ge$ 1 $b$-tagged jet:

\begin{itemize}[label={-}]
\item $\Delta R ($large-R jet, b-tagged jet$) < $ 1 or,
\item the small-R jet is $b$-tagged.
\end{itemize}}\\
\hline\hline
\end{tabular}

%}
\caption{Event selections applied in the \emph{Resolved} and \emph{Boosted} analyses.}
\label{tab:cuts}
\end{table}

For the selected events the $\ttb$ system is reconstructed based on the event topology:
\begin{itemize}
\item \textbf{Resolved analysis:} The leptonic top is reconstructed using the $b$-tagged jet nearest in $\Delta R$ to the lepton
  and the missing-momentum four vector, the hadronic top is reconstructed using the other $b$-tagged jet and the
  two light jets with invariant mass closest to the $W$ mass.
\item \textbf{Boosted analysis:} The leptonic top is reconstructed using the \emph{lep-jet} candidate, the lepton
  and the missing-momentum four vector, the \emph{had-jet} candidate is directly considered as the hadronic top.
\end{itemize}

In order to validate our simulations of SM top-quark pair-production we compare our predictions
against ATLAS data for the \emph{Boosted} and \emph{Resolved} selection, supplemented by studies of systematic
variations. To begin with, we check the impact of the grooming procedure on the reconstructed hadronic-top
candidate mass, \emph{i.e.} the mass of the \emph{had-jet} candidate in the \emph{Boosted} event selection.
We consider event samples from \Sherpa\ and \mg5, based on the leading-order matrix element for top-quark
pair production, labelled as $0j$. In these calculations, \emph{i.e.} without merging-in higher-multiplicity matrix elements,
we set the renormalization ($\mu_R$) and factorization scale ($\mu_F$) to 
\begin{equation}
\mu_R^2=\mu_F^2 = \frac14 [ m_t^2 + \frac12 (p_{T,t}^2 + p_{T,\bar{t}}^2) ]\,,
\end{equation}
with $p_{T,t}$ ($p_{T,\bar{t}}$) the transverse momentum of the decaying (anti) top quark.

\begin{figure}[h!]
\includegraphics[width=0.49\textwidth]{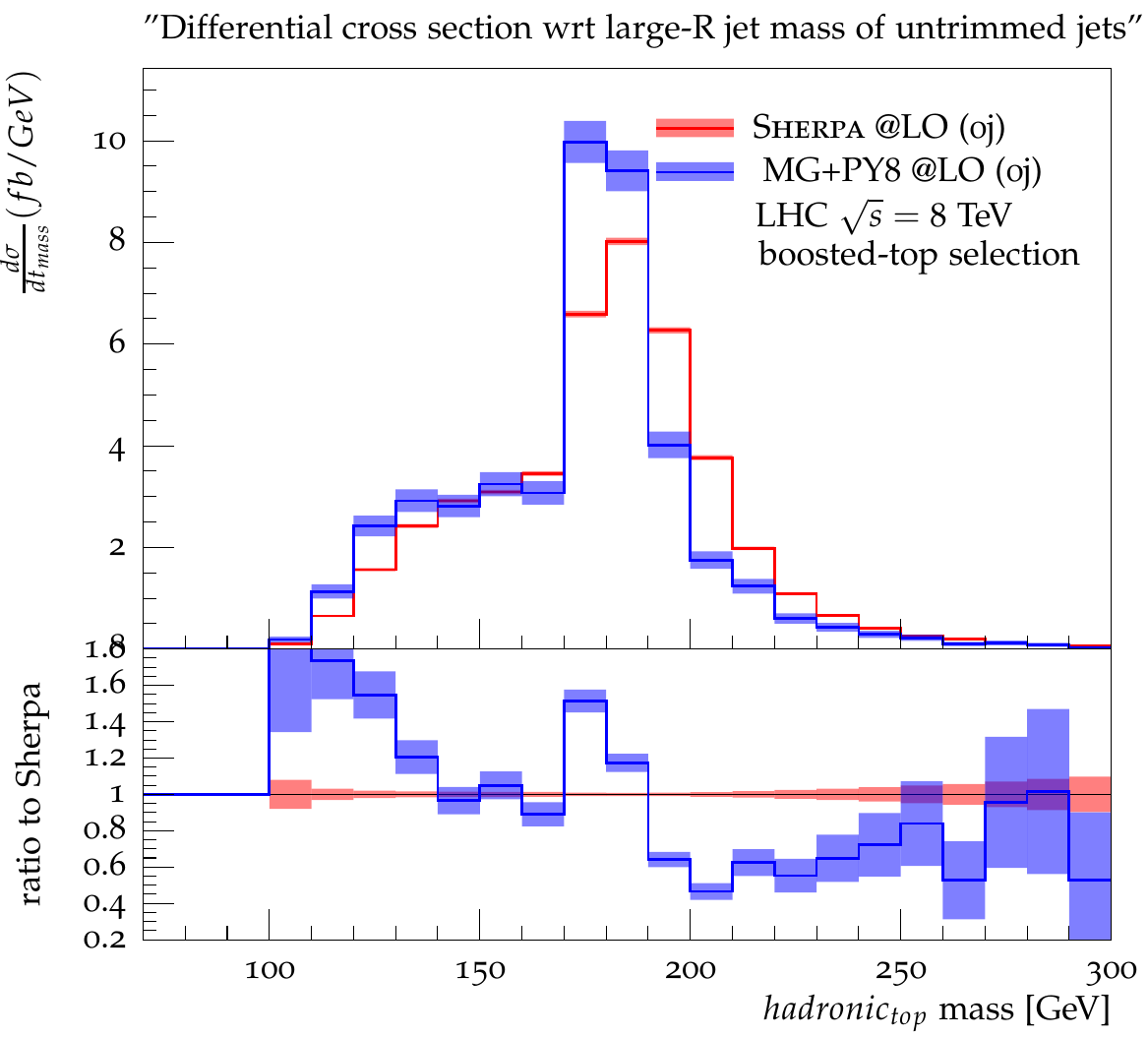}
\includegraphics[width=0.49\textwidth]{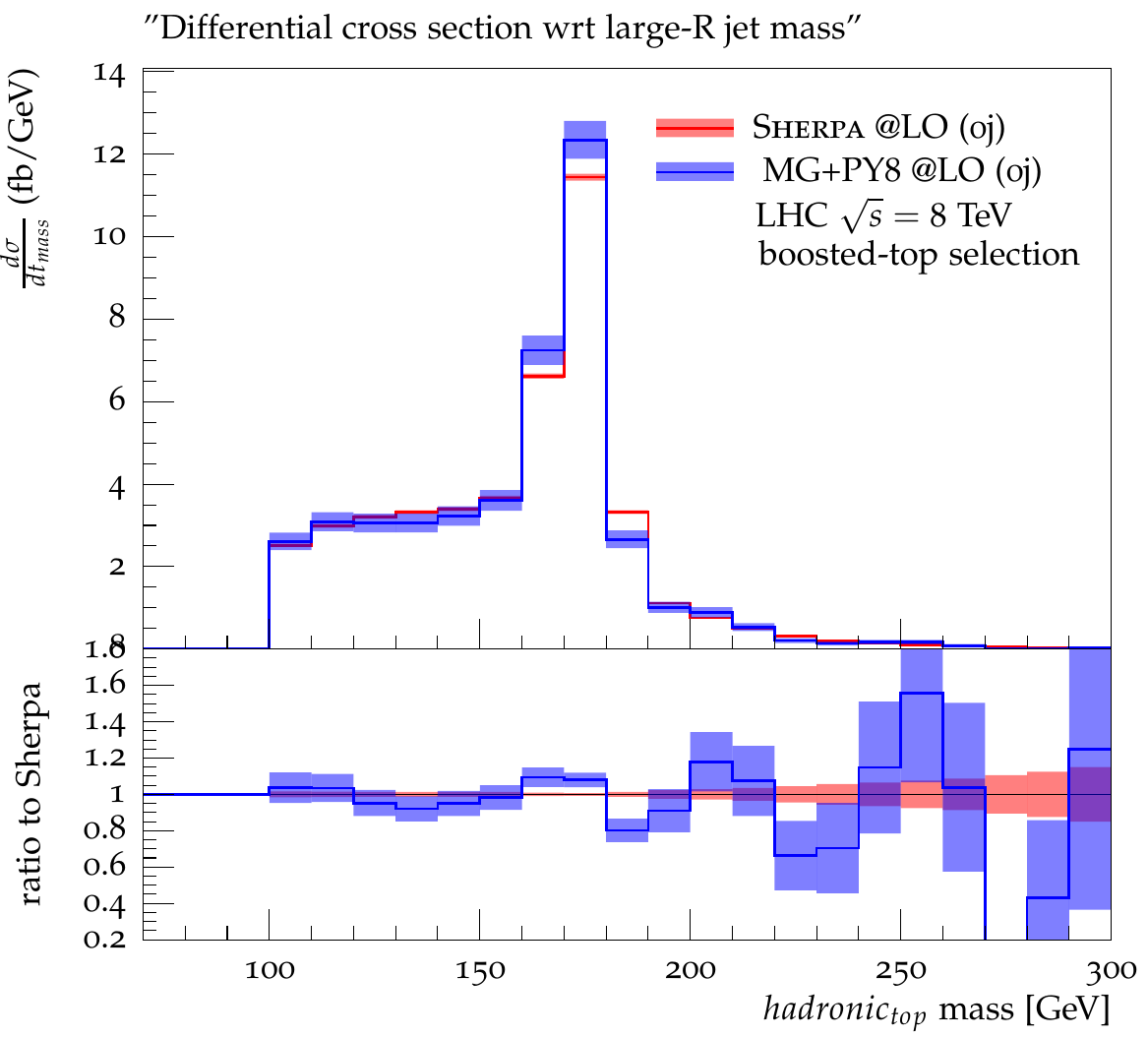}
\caption{Invariant mass distribution of the hadronic-top candidates in the \emph{Boosted} event selection.
  The theoretical predictions from \Sherpa\ and \mg5+\textsc{Pythia}8 are based on LO matrix elements
  dressed with parton showers, left panel without and right panel with applying the trimming procedure.}
\label{fig:Trimming}
\end{figure}

In Fig.~\ref{fig:Trimming} we present the resulting invariant-mass distributions obtained from \Sherpa\ and \mg5
before and after applying the grooming procedure. Comparing the untrimmed distributions (left panel) both samples
exhibit a clear peak at the nominal top-quark mass. However, due to parton-shower radiation and non-perturbative
corrections from hadronization and underlying event the peak is rather broad and sizeable differences are observed
when comparing the predictions from \Sherpa\ and \mg5+\textsc{Pythia}8. Note that the uncertainty bands shown represent
the statistical uncertainty of the samples only. When applying the trimming procedure to the \emph{had-jet} candidates
the mass distributions agree to a much better degree, both in the tails of the distribution and the peak region. Therefore,
trimming of the large-R jets significantly reduces the dependence on the generator and the details of its parton-shower
formalism and the modelling of non-perturbative effects. 

In Figs.~\ref{fig:scaleUncertaintyLO} and \ref{fig:scaleUncertaintyNLO} we compare predictions from \Sherpa\ based on
LO and NLO matrix elements against data measured by the ATLAS experiment for the \emph{Boosted} (left panels) and the
\emph{Resolved} (right panels) event selections. For the MEPS@LO sample we merge LO QCD matrix elements for
$t\bar{t}+0,1,2,3$jet production dressed with the \Sherpa\ dipole parton shower~\cite{Hoeche:2009rj}. The merging-scale
parameter is set to $Q_{\rm cut}=20\,{\rm GeV}$. The MEPS@NLO sample combines QCD matrix elements at NLO for
$t\bar{t}+0,1$jet and $t\bar{t}+2,3$jets at LO according to the methods described in
\cite{Hoche:2010kg,Hoeche:2012yf}, again using a merging scale of $Q_{\rm cut}=20\,{\rm GeV}$. Both methods share
the event-wise reconstruction of an underlying $jj\to t\bar t$ core process through consecutive clusterings of the
external legs. For this reconstructed core process the renormalization and factorization scales are set
to $\mu_R=\mu_F=\mu_{\rm core}$, with
\begin{equation}
\mu^2_{\rm core} = \frac14 [ m_t^2 + \frac12 (p_{T,t}^2 + p_{T,\bar{t}}^2) ]\,.
\end{equation}
For the reconstructed clusterings the strong coupling is evaluated at the respective splitting scale. The
scale $\mu_{\rm core}$ is furthermore used as the resummation, \emph{i.e.} parton-shower starting scale, denoted $\mu_Q$. To
assess the scale uncertainty of the predictions we perform variations by common factors of $2$ and $1/2$ for
the core scale and the local splitting scales, using the event-reweighting technique described in
\cite{Bothmann:2016nao}. In the figures the resulting uncertainty estimate is represented by the red band, while
the blue band indicates the statistical uncertainty.

\begin{figure}[h!]
\includegraphics[width=0.49\textwidth]{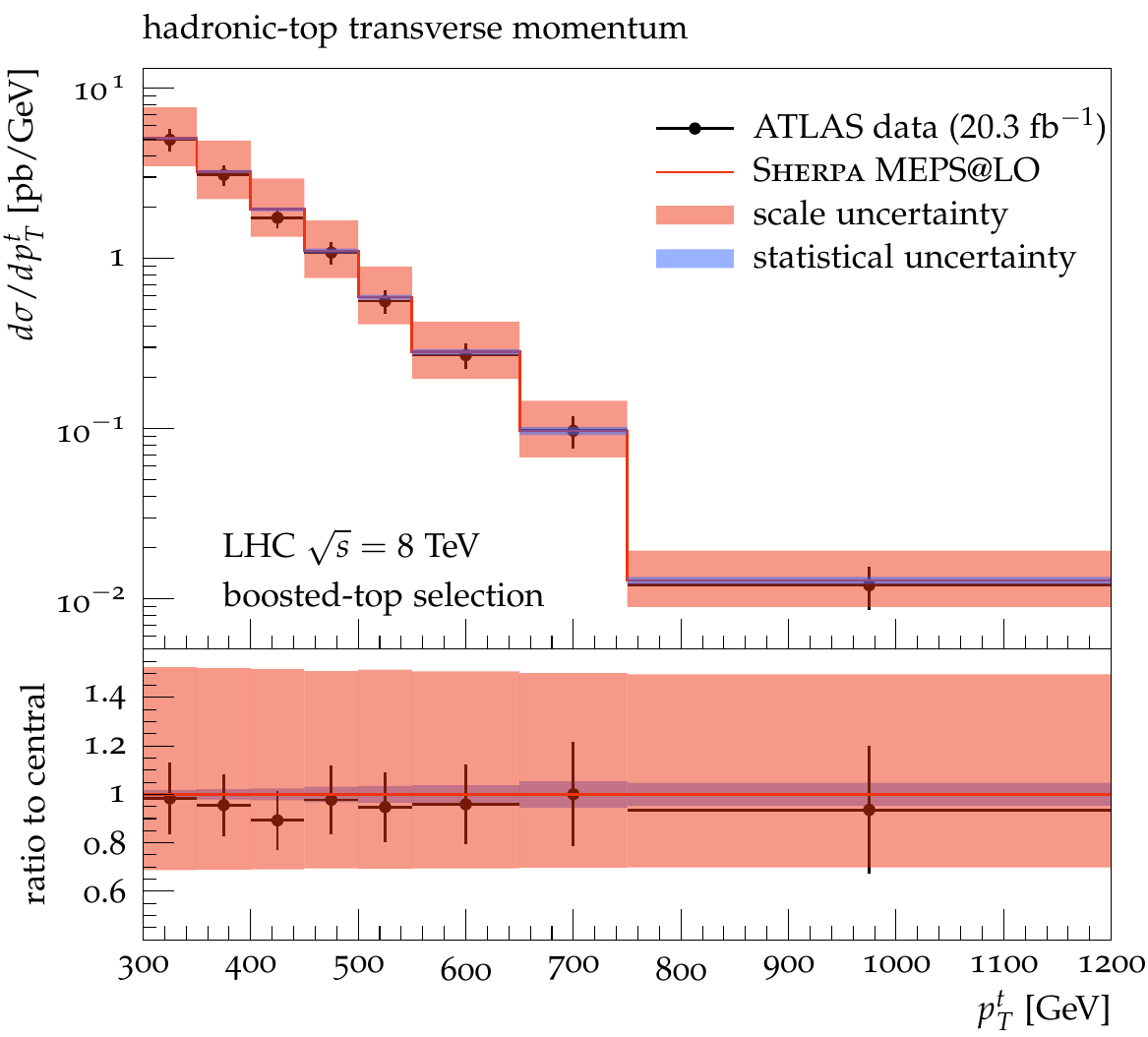}
\includegraphics[width=0.49\textwidth]{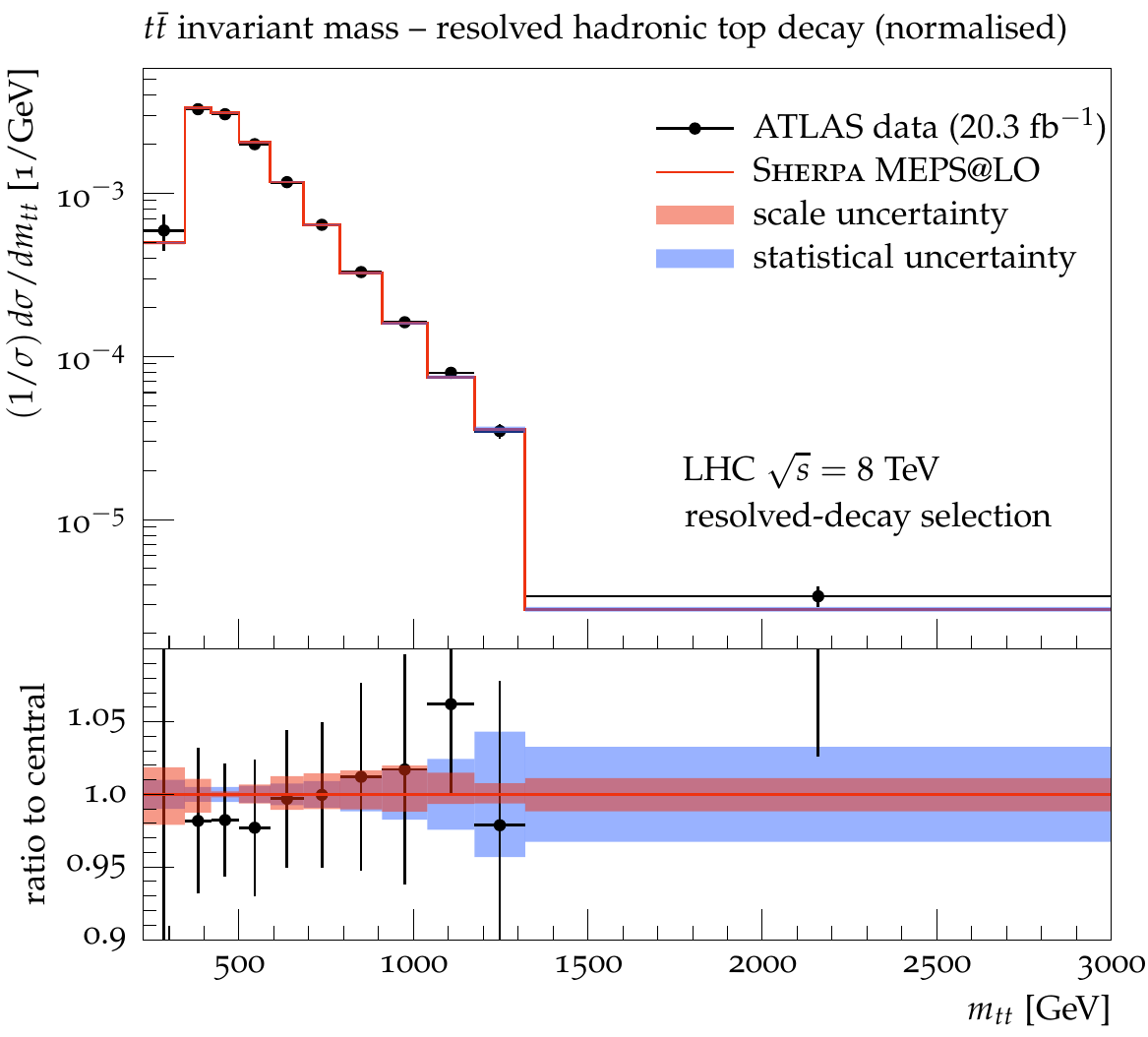}
\caption{Comparison of predictions based on \Sherpa\ MEPS@LO simulations to data measured
  by the ATLAS experiment. The left panel shows the $p_T$ of the hadronic top in the \emph{Boosted} selection,
  data taken from ~\cite{Aad:2015hna}.
  In the right panel the reconstructed invariant mass of the $\ttb$ system in the \emph{Resolved} event selection is
  depicted, with data taken from~\cite{Aad:2015mbv}.}
\label{fig:scaleUncertaintyLO}
\end{figure}

\begin{figure}[h!]
\includegraphics[width=0.49\textwidth]{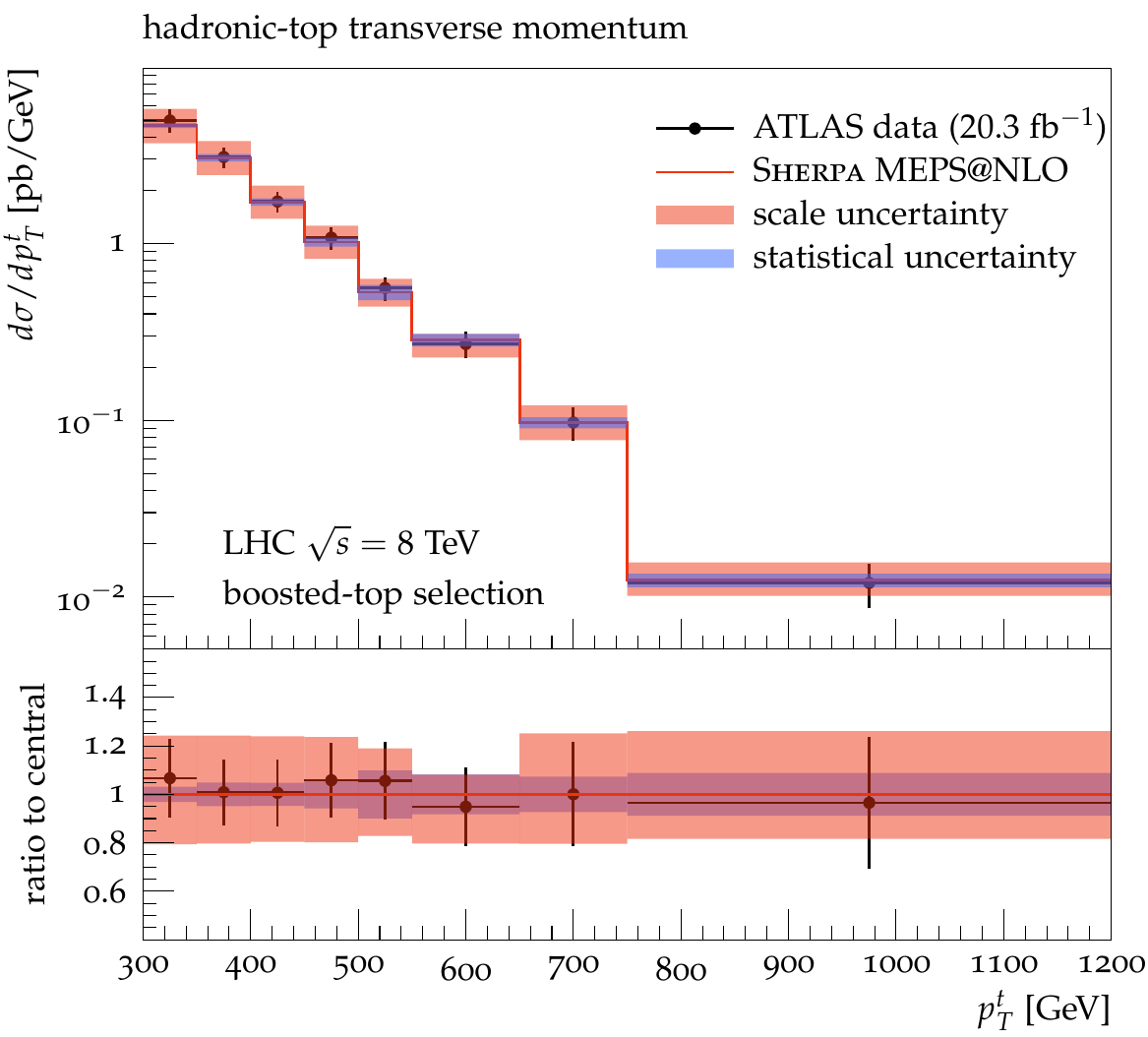}
\includegraphics[width=0.49\textwidth]{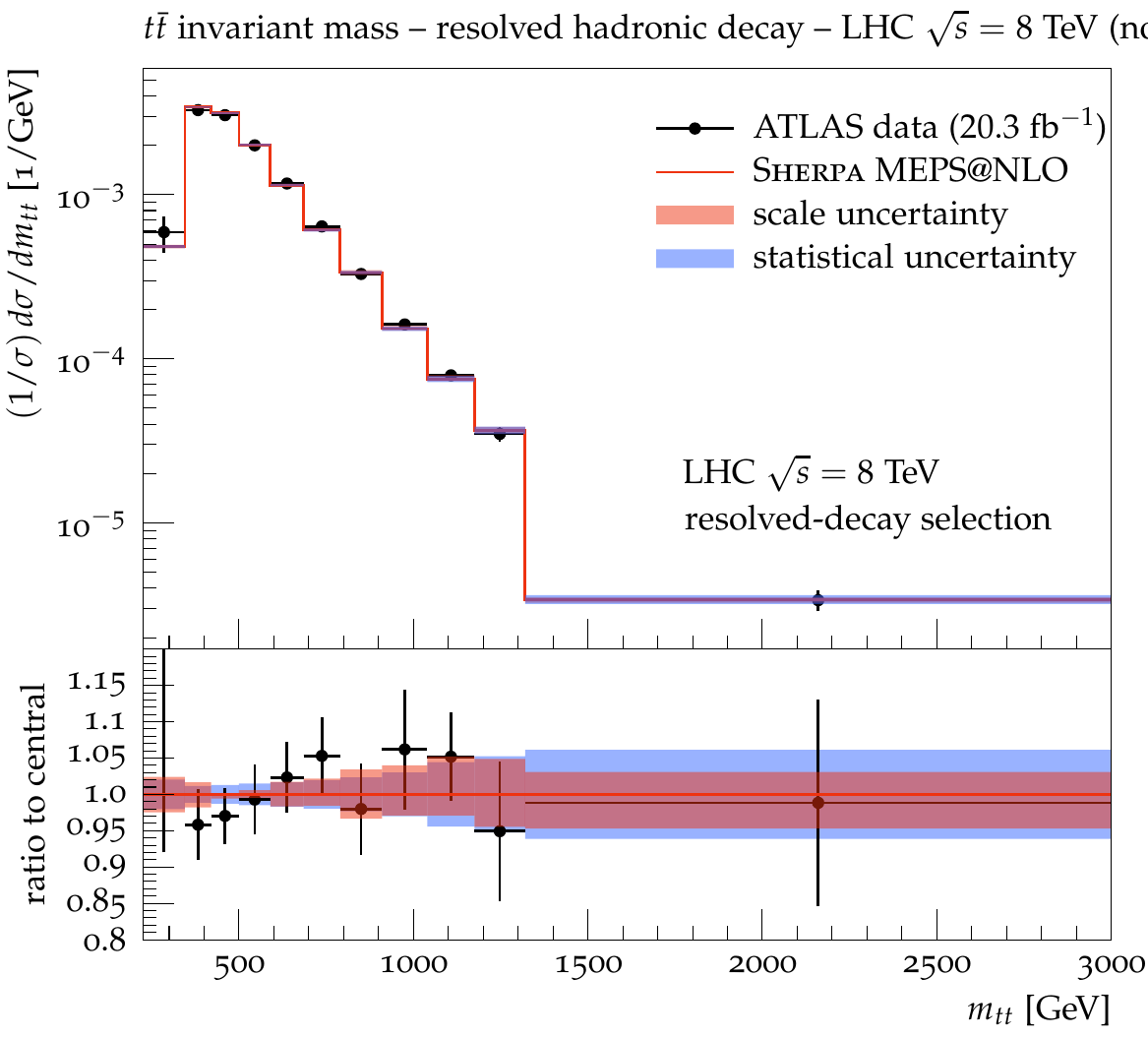}
\caption{As Fig.~\ref{fig:scaleUncertaintyLO} but based on \Sherpa\ MEPS@NLO simulations.}
\label{fig:scaleUncertaintyNLO}
\end{figure}

For the boosted-top selection we show the transverse-momentum distribution of the hadronic-top candidate in the left
panels of Figs.~\ref{fig:scaleUncertaintyLO} and \ref{fig:scaleUncertaintyNLO}, respectively. Notably, both samples,
\emph{i.e.} the MEPS@LO and the MEPS@NLO prediction, describe the ATLAS measurement \cite{Aad:2015hna} very well, both in
terms of the production rate and in particular concerning the shape of the distribution. For the MEPS@LO result the
scale uncertainty is quite significant, reaching up to 50\%. However, the dominant effect is a mere rescaling of the
total production rate, the shape of the distribution stays almost unaltered. This is also observed for the MEPS@NLO
sample, however, the scale uncertainty reduces to $\pm20\%$.

For the resolved-decay selection we compare the \Sherpa\ MEPS@(N)LO predictions for the reconstructed invariant mass
of the $t\bar t$ system against data from the ATLAS experiment~\cite{Aad:2015mbv}, see right panels of
Figs.~\ref{fig:scaleUncertaintyLO} and \ref{fig:scaleUncertaintyNLO}. Note that the data and the theoretical predictions
are normalized to their respective fiducial cross section. The MEPS@LO and MEPS@NLO results agree very well with the data.
For this normalized distribution the scale uncertainties largely cancel. For the MEPS@LO sample this results in an
uncertainty estimate of $\pm 2\%$. For the MEPS@NLO sample the shape modifications induced by the scale variations
amount to $\pm 5\%$.

For both observables considered, the MEPS@(N)LO predictions from \Sherpa\ yield a very satisfactory description of
the data. No significant alteration of the distributions shape is observed upon inclusion of the QCD one-loop
corrections in the MEPS@NLO sample. However, in particular the uncertainty on the production rate reduces significantly.
For the normalized top-pair invariant mass distribution we consider the more realistic $\pm 5\%$ estimate from the
MEPS@NLO calculation. By normalizing the distribution to the cross section in a certain mass window, this
uncertainty might in fact be reduced further, cf. Ref.~\cite{Czakon:2016vfr}, where, ultimately, an uncertainty
estimate of ${\cal{O}}(1\%)$ was quoted for the corresponding NNLO QCD prediction. 

In what follows we want to study the imprint of New Physics resonant contributions on the top-pair invariant
mass distribution. To this end we currently rely on a leading-order description of the signal, interfering with the
corresponding SM amplitudes. However, from the considerations above we can conclude that the MEPS@LO calculation
of the SM production process captures the dominant QCD corrections, which are of real-radiation type. To illustrate
this further, we present in Fig.~\ref{fig:NLOxLO} a comparison of MEPS@LO samples using different parton-multiplicity
matrix elements for the mass and the transverse momentum of the $t\bar{t}$ system in the \emph{Boosted} selection.
These results get compared to the corresponding MEPS@NLO prediction described above. 

\begin{figure}[h!]
\includegraphics[width=0.49\textwidth]{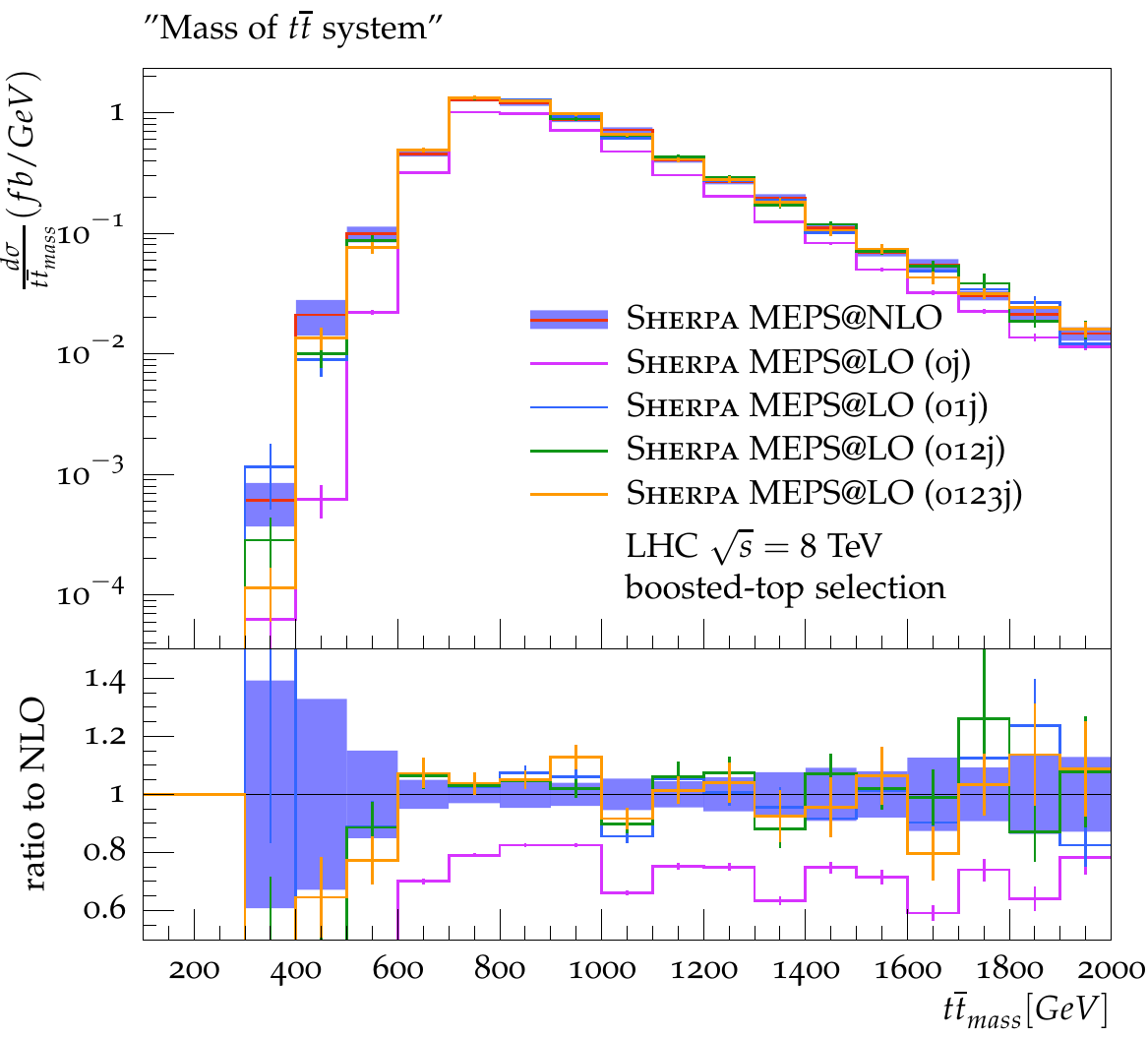}
\includegraphics[width=0.49\textwidth]{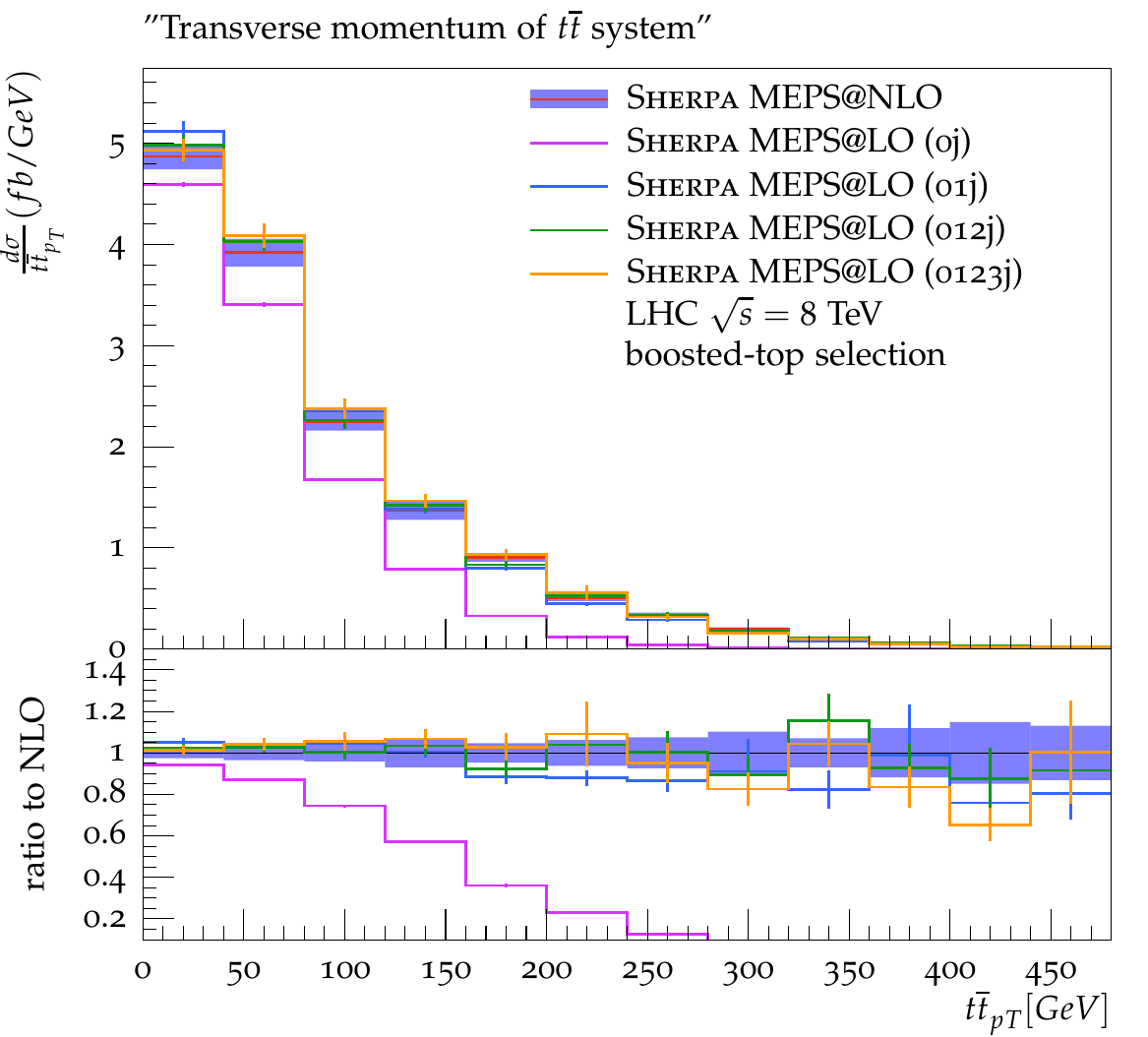}
\caption{Comparison of MEPS@LO predictions based on different maximal parton-multiplicity matrix elements and the
  MEPS@NLO calculation for the \emph{Boosted} event selection. The left panel shows the top-pair invariant mass,
  the right panel the transverse momentum of the $\ttb$ system. }
\label{fig:NLOxLO}
\end{figure}

For the top-pair invariant mass all the predictions with at least one extra hard jet agree within their statistical errors. In particular, even the
MEPS@LO sample, based on merging the LO matrix elements for $t\bar{t}+0,1$jet only, well reproduces the MEPS@NLO result and greatly improves the 0jet sample. 
As might be expected, for the transverse momentum of the $t\bar{t}$ system, the inclusion of higher-multiplicity
matrix elements improves the agreement with the MEPS@NLO result. The MEPS@LO calculation based on $t\bar{t}+0,1$jet
predicts a somewhat softer spectrum, \emph{i.e.} is lacking configuration corresponding to multiple hard emissions.
However, the bulk of the events in the \emph{Boosted} selection is reasonably modeled by this simple LO merging
setup and describes the data presented above very well. We will therefore rely on this setup when invoking New
Physics contributions. 

In the following we also introduce a simple \emph{Parton Analysis}, used to quantify the effect of the NP without
any smearing due to the reconstruction of the top quarks. In the \emph{Parton Analysis} no cuts are applied to the
events and the two top quarks are identified, before any decay, using truth-level information from the generator.
%Since the top quarks are identified using generation information it is impossible to compare the \emph{Parton Analysis}
%distributions with data.

\section{Simplified model}
\label{sec:model}

Several models of NP predict resonances decaying to top-quarks. Scalar resonances in particular have
large branching ratios in this decay channel due to the fact that their couplings with fermions are often
proportional to the fermion masses. In this case, the resonance is at the LHC dominantly produced via
gluon fusion through loops of colored particles. These colored particles can be either light compared
to the resonance (like the top quark itself), in which case the structure of the loop is resolved as
illustrated in Fig.~\ref{fig:diag}(a), or they can be heavy, in which case a point-like interaction
sketched in Fig.~\ref{fig:diag}(b) can describe the interactions. 
%
%via new heavy physics which generates a point-like interaction, either heavy colored particles or anomaly induced couplings (illustrated in \fig{fig:diag}(b)). 

\begin{figure}
%\begin{tabular}{c}
\includegraphics[width=0.8\textwidth]{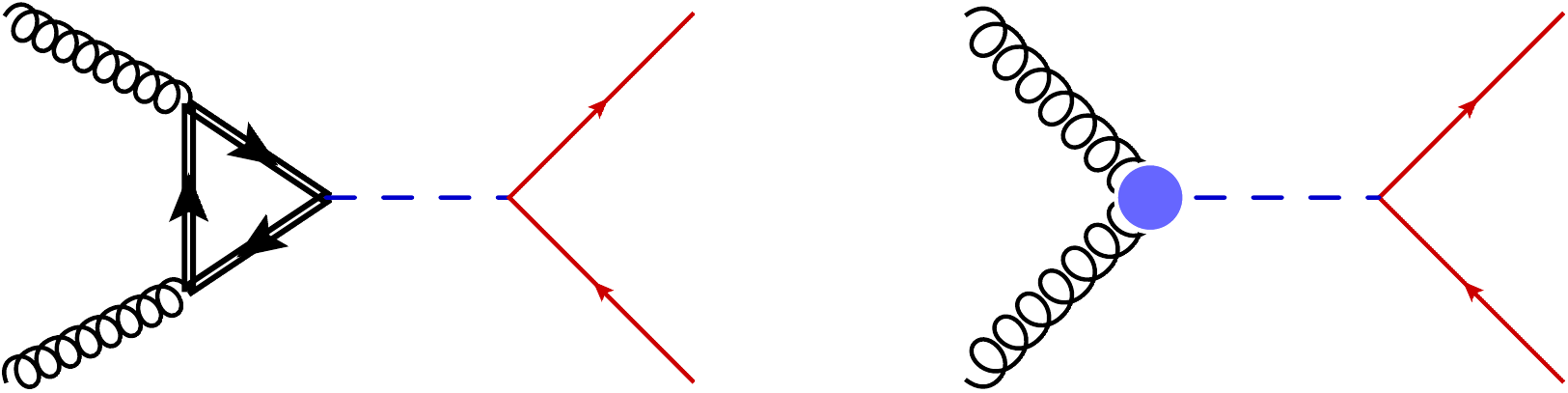} \\
(a) \hspace{6.5cm} (b) 
%\end{tabular}
\caption{Feynman diagrams for production of a scalar resonance with subsequent decay into top-quarks, mediated by a \emph{resolved}
  loop (a) or via high-scale New Physics (b).}
\label{fig:diag}
\end{figure}

It has been shown in~\cite{Manohar:2006ga} that the most general scalar extension of the SM which couples to fermions and maintains
naturally small flavour changing neutral currents is provided by scalars with the same quantum numbers of the Higgs doublet
or that transform as a color octet $(\bf 8, 2)_{1/2}$ under the SU(3)$\times$SU(2)$\times$U(1) SM gauge group.
Color neutral and octet scalars arise also naturally in several models of dynamical EWSB, such as in the seminal
Farhi-Susskind model~\cite{Farhi:1980xs} and models where the top is partially composite~\cite{Belyaev:2016ftv}.
Although the specific origin of the scalar-top couplings is important, determining  the relation to other couplings and their magnitudes, we here adopt a more  phenomenological simplified approach relevant for top-quark pair production, in which the \emph{left-handed} top is stripped off from its doublet and couples directly to the scalars. 

%Both CP-even and CP-odd states may have relevant contribution in $\ttb$ final state. In the color neutral case, the CP-odd states is more likely to decay into top-quarks, since its decay into weak bosons is suppressed. The color octet case is more democratic since both CP-even and CP-odd states must decay into colored final states.

In our simplified model we assume the only \emph{light} state \emph{running} in the loop to be the top-quark. 
This is a good approximation if two conditions are fulfilled: (i) - the bottom-quark contribution is suppressed;
and (ii) - the extra states contributing significantly to the gluon--scalar couplings are heavy
(at least as much as the scalar resonance itself). This is a good approximation in many models beyond the SM.
In the THDM~\cite{Branco:2011iw} for example, there is no new particle living at higher scale apart from the new
scalar sector. Moreover, the loop of bottom-quarks is usually suppressed in the cases relevant for $\ttb$ production.
%, \emph{e.g.} in Type-I and Type-II models with small $\tan\beta$. 
Specializations of the THDM such as the Minimal Supersymmetric Standard Model (MSSM) where the super-partners are heavy enough to be integrated out can also be described in this framework. 
Composite models typically predict relatively degenerate spectra of first excitations, thus they can be usually described by the effective point-like interaction.  
% and some realizations of bosonic Technicolor~\cite{Carone:2012cd} are also well approximated by this hypothesis, since the composite states have similar masses.
Similarly, for the color octet in the model of Manohar and Wise~\cite{Manohar:2006ga} the scalars are produced purely by top and bottom loops. 
In some other models intermediate states much lighter than the first scalar excitations are present, \emph{e.g.} top partners and stops may be light in some models of partial compositeness and SUSY -- in these cases our approximation is not applicable.

Under this assumption we can describe the scalar sector interactions relevant for $t\bar{t}$ production via the following Lagrangian:
\begin{eqnarray}
\mathcal{L}_\phi &=&  i c^\eta_t\frac{m_t}{v} \bar{t}\gamma_5 t \eta +  c^\sigma_t \frac{m_t}{v} \bar{t} t \sigma 
             + i c^{\tilde{\eta}}_t\frac{m_t}{v} \bar{t}\gamma_5 \frac{\lambda^a}{2} t \tilde{\eta}^a 
  + c^{\tilde{\sigma}}_t\frac{m_t}{v} \bar{t} \frac{\lambda^a}{2} t {\tilde{\sigma}}^a \nonumber \\
  %\label{eq:ytt}\\
       &+& c_g^\sigma \frac{\alpha_S}{12\pi v}\sigma G^a_{\mu\nu}G^{a\mu\nu} 
        - c_g^\eta \frac{\alpha_S}{8\pi v}\eta G^a_{\mu\nu}\widetilde{G}^{a\mu\nu} \nonumber \\
        &-& c_g^{\tilde{\eta}} \frac{\alpha_S}{8\pi v} \tilde{\eta}^a d^{abc} \widetilde{G}^{a\mu\nu} G^{b\rho\sigma} 
        + c_g^{\tilde{\sigma}} \frac{\alpha_S}{12\pi v}{\tilde{\sigma}}^a d^{abc} \widetilde{G}^{a\mu\nu} G^{b\rho\sigma} \,.        
\label{eq:ygg}
\end{eqnarray}
It contains a CP-odd isosinglet scalar $\eta$, a CP-even isosinglet scalar $\sigma$, a  CP-odd color octet scalar $\tilde{\eta}$ and a CP-even octet scalar $\tilde{\sigma}$ which we collectively call $\phi$.
 $G^{\mu\nu}$ is the gluon field-strength tensor, $\widetilde{G}^{\mu\nu}=\frac{1}{2}\epsilon^{\mu\nu\rho\sigma}G_{\rho\sigma}$, $\lambda^a$ are the SU(3) generators and $d^{abc}=\frac{1}{4}Tr[\lambda^a{\lambda^b,\lambda^c}]$ is the fully symmetric SU(3) tensor.

The top-quark loops generate form factors that describe the gluon-scalar interaction.
The loop triangles contribute to the trilinear $gg\phi$ vertices in the form
 \begin{eqnarray}
\eta g_a^\mu(k_1) g_b^\nu(k_2) &:&\quad \frac{\alpha_S c_t^\eta}{2\pi v} A^{A}_{1/2}\left(\frac{s}{4m_t^2}\right) 
   \epsilon^{\mu\nu\lambda\sigma}k_{1\lambda} k_{2\sigma}\delta_{ab} \,,\\
\sigma g_a^\mu(k_1) g_b^\nu(k_2) &:&\quad \frac{\alpha_S c_t^\sigma}{3\pi v} A^{S}_{1/2}\left(\frac{s}{4m_t^2}\right) 
 (k_1^\nu k_2^\mu-k_1\cdot k_2 g^{\mu\nu})\delta_{ab} \,,\label{eq:formfactors}
\end{eqnarray}
with
\begin{eqnarray}
A^{A}_{1/2}(\tau)&=&f(\tau)/\tau \,, \label{eq:AA}\\
A^{S}_{1/2}(\tau)&=&\frac{3}{2\tau^2}\left(\tau+(\tau-1)f(\tau)\right) \,, \label{eq:AS}\\
f(\tau) &=& 
    \begin{cases}
      \text{arcsin}^2(\sqrt{\tau}), & \quad \tau\leq 1 \\
      -\frac{1}{4}\left[\log\left(\frac{1+\sqrt{1-\tau^{-1}}}{1-\sqrt{1-\tau^{-1}}} \right)-i\pi \right]^2, & \quad \tau>1\,.
    \end{cases}
\label{eq:ftau}
\end{eqnarray}  
Similar expressions for the color octet top-quark loop generated form factor can be found \emph{e.g.} in~\cite{Hayreter:2017wra}. 

As a matter of fact, resonant top pair production is accompanied by other signatures. In particular, diphoton,
  dijet, $\gamma Z$, $ZZ$ and $W^+W^-$ signatures are generated via diagrams induced by a top-quark loop, and in
  general by high-scale physics. Tree-level $ZZ$, $W^+ W^-$ decay channels are typically present for a scalar state,
  while decays into lighter fermions are typically suppressed.  Color octets decays into $g\gamma$ and $g Z$ might
  give striking signatures. The detailed analysis of these channels is not in the scope of this work, however, we
  provide some qualitative discussion about the regions in parameter space where they can be competitive in
  sensitivity to $t\bar{t}$ search.

  Loop (or anomaly) induced decays are typically suppressed and might be competitive to $t\bar{t}$ searches only
  for small Yukawa couplings $c_t$. They are often the only possible decay channels for pseudo-scalars besides
  that into $t\bar{t}$. As an example, consider some partial widths of a color-singlet pseudo-scalar 
\begin{eqnarray}
\Gamma_{\eta\to t\bar{t}} &=& \frac{3}{8\pi} %y_{\eta t}^2
    \frac{m_t^2}{v^2}(c^\eta_t)^2\,  m_{\eta}\sqrt{1-4 m_t^2/m_{\eta}^2}  \,, \\
\Gamma_{\eta \to gg} &\simeq&  \frac{\alpha_s^2m_{\eta_X}^3}{32 \pi^3 v^2} 
      \left|  c_t^\eta A^{A}_{1/2}\left(\frac{m_\eta^2}{4m_t^2}\right)+ c_g^\eta \right|^2 \,,\\
      \label{eq:gamma-eta-gg}
\Gamma_{\eta \to \gamma\gamma} &\simeq&  \frac{\alpha^2 m_{\eta_X}^3}{256 \pi^3 v^2} 
      \left|  c_t^\eta 3(2/3)^2 A^{A}_{1/2}\left(\frac{m_\eta^2}{4m_t^2}\right)+ c_\gamma^\eta \right|^2\,.
%\Gamma_{\eta \to \gamma\gamma} &\simeq&  \frac{\alpha^2 m_{\eta_X}^3}{256 \pi^3 v^2} 
%     \left|  c_W^\eta+ c_B^\eta \right|^2 \\
%%      
%\Gamma_{\sigma\to t\bar{t}} &=& \frac{3}{8\pi} 
%    \frac{m_t^2}{v^2}(c^\sigma_t)^2\,  m_{\sigma}\left(1-m_t^2/m_{\sigma}^2\right)^{3/2}  \,, \\
%\Gamma_{\sigma \to gg} &\simeq&  \frac{\alpha_s^2 m_{\sigma}^3}{72 \pi^3 v^2} 
%      \left|  c_t^\sigma A^{S}_{1/2}\left(\frac{m_\sigma^2}{4m_t^2}\right)+ c_g^\sigma \right|^2
\label{Eq:partialwidths}
\end{eqnarray}
Here we parametrize the photon interaction with $\eta$ by the following gauge invariant operators
\begin{equation}
\mathcal{L}_{\phi,\gamma} = - c_W^\eta \frac{\alpha}{8\pi v}\eta W^i_{\mu\nu}\widetilde{W}^{i\mu\nu} - c_B^\eta \frac{\alpha}{8\pi v}\eta B_{\mu\nu}\widetilde{B}^{\mu\nu}\,,
\end{equation}
with $c_\gamma^\eta\equiv c_W^\eta+c_B^\eta$.
These operators also give rise to decays into weak bosons, but not competitive in
sensitivity to diphoton searches (unless there is some cancellation in $c_W+c_B$).
From the above expressions it can be noticed that the $gg$ partial width is much larger than $\gamma\gamma$,
however, the corresponding search is not as competitive to the diphoton channel due to
the clean signature of the latter. 

On the other hand, scalar resonances tend to decay into weak bosons at tree level, with
large contributions to their decay width and good sensitivity in the corresponding channels. 

The color octets have more unexplored signatures, like \emph{e.g.} $g\gamma$, studied for example in Ref.~\cite{Aad:2015ywd,Aaboud:2017nak}.

%%%%%%%%%%%%%%%%%%%%%%%%%%%%%%%%%%%%%%%%%%

\subsection{Model description and simulation}

Our goal is to achieve accurate predictions for a wide parameter range of our generic model in an efficient and fast way. 
For this purpose, the Lagrangian given in \eq{eq:ygg} has been implemented into the \Feynrules~\cite{Alloul:2013bka} package to produce
a corresponding UFO model file~\cite{Degrande:2011ua}. 
The required helicity amplitudes have been extracted to C++ codes via the Madgraph~\cite{Alwall:2011uj} program and incorporated in the \Rivet\,   analyses in order to perform a \emph{reweighting} method and reproduce the signal line-shape.
% for a large region of parameter space without much computational cost. 
To this end, each event of the \Sherpa\ SM event sample is given a weight, $w$, proportional to the ratio of the amplitudes,
\begin{equation}
w=\frac{\overline{|\mathcal{M}_{\rm SM}+\mathcal{M}_\phi|^2}}{\overline{|\mathcal{M}_{\rm SM}|^2}},
\end{equation}
where $\overline{|\mathcal{M}_{\rm SM}|^2}$ is the SM amplitude  squared summed and averaged over color and spin. In the numerator the amplitude $\mathcal{M}_\phi$ corresponding to the resonant diagrams depicted in Fig.~\ref{fig:diag} is added on top of the SM diagrams. The further decay of top quarks is included neglecting non-resonant diagrams. Therefore, the full  process in \eq{eq:process} -- including possible extra hard radiation -- is considered with full spin correlation of the top-quark decays. 

We note that our signal includes not only the purely resonant contribution. The complete squared amplitude can be split into three contributions:
\begin{equation}
|\mathcal{M}_{\rm SM}+\mathcal{M}_\phi|^2= 
|\mathcal{M}_{\rm SM}|^2+|\mathcal{M}_{\phi}|^2
+2{\rm Re}\, \mathcal{M}_{\rm SM}^*\mathcal{M}_{\phi}
\equiv B_\mathcal{M}+S_\mathcal{M}+I_\mathcal{M}\,.
\end{equation}
The last term defines the SM background ($B_\mathcal{M}$), the pure signal ($S_\mathcal{M}$) and the interference between signal and SM ($I_\mathcal{M}$). 
%giving origin to pure resonant contribution and the interference between the new physics signal and the SM continuum. 

We use as the test observable the 
$m(t\bar{t})$ distribution of the signal hypothesis $H$ normalized bin-by-bin to the SM QCD prediction,
\begin{equation}
r(H)\equiv\frac{d\sigma_H/dm}{d\sigma_{\rm SM}/dm}\,.
\label{eq:r}
\end{equation}
The signal hypothesis differential cross section $d\sigma_H/dm$ is defined as the total differential cross section subtracted by the SM prediction. 
Such normalized distribution is less affected by systematic errors, \emph{i.e.} theoretical uncertainties~\cite{Czakon:2017wor}. 

In order to assess the importance of the interference we study both the full signal including interference $d\sigma_{S+I}/dm$ and the pure signal hypothesis neglecting interference $d\sigma_S/dm$.
To simplify the notation in the remaining of the text we use the following definitions:
\begin{equation}
d\sigma_S/dm\equiv S_\sigma,\quad d\sigma_I/dm\equiv I_\sigma, \quad d\sigma_{\rm SM}/dm\equiv B_\sigma\,.
\end{equation}
Interference between signal (Fig.~\ref{fig:diag})  and QCD diagrams are known to be important in this process.  In fact,
they can completely change the line-shape of the resonance from a pure Breit-Wigner peak to a peak-dip structure, or
even dip-peak, pure dip or an enhanced peak~\cite{Gaemers:1984sj,Dicus:1994bm,Frederix:2007gi,Gori:2016zto,Jung:2015gta,Djouadi:2016ack}. QCD corrections to this effect have recently been computed~\cite{Bernreuther:2015fts,Hespel:2016qaf,BuarqueFranzosi:2017jrj} and shown to be important.
A pilot experimental analysis investigating such interference effects has been presented recently~\cite{Aaboud:2017hnm}.

The form factors in \eq{eq:formfactors} have been implemented in the helicity amplitudes used in the reweighting step.
However, the corresponding box diagram contributing to the four-gluon--scalar coupling was kept as an effective vertex without
momentum dependence. For the color octet the form factor is approximated by a fixed momentum flowing through the loop
that is equal to the mass of the resonance. The interference between top-quark loops and point-like interactions is also
manifest in the calculation.

Higher-order QCD corrections are partially taken into account through the radiation of extra gluons in the MEPS@LO simulation.
The contribution from real-emission $t\bar{t}j$ matrix elements also get reweighted with the NP theory hypotheses.
 
We note however that the method neglects the signals' color-singlet
color flow contribution when attaching parton showers, which affects the subsequent radiation pattern only. 
We nevertheless found that these effects are small in the description of the top-pair mass distribution.
In Fig.~\ref{fig:rwgtXfull} we show the distribution of variable $r(S)$ defined in \eq{eq:r} for a color-singlet
pseudo-scalar of mass $1.5$ TeV in the pure signal hypothesis, comparing the \Sherpa ~\emph{reweighted} events with a dedicated simulation of the full
process with \mg5+\textsc{Pythia}8. 
In the latter, the color-flow contribution corresponding to the signal diagrams are considered as
seeds for the subsequent parton shower. The error near the resonance peak is
about 10\% and the reweighted prediction underestimates the yields. 
We removed the top-quark loop form factor considering only the effective scalar-gluon coupling for this
comparison. The distributions were derived according to the \emph{Parton Analysis} framework
described in \sec{sec:analysis}. In the more realistic boosted analysis we expect the reweighting
method to predict a more smeared distribution due to the extra connected color lines that favor
extra hard radiation connecting the top quarks with initial gluons.  We will neglect these effects and employ the \emph{reweighting}
method in what follows to make predictions for a large region of parameter space of the model, while avoiding massive
time and machine consuming event generation and ``fake" MC statistical error. 
Our results are expected to give conservative limits since for colour-singlet resonances the signal color flow induces
less smearing of the resonance peak.

\begin{figure}
\includegraphics[width=0.49\textwidth]{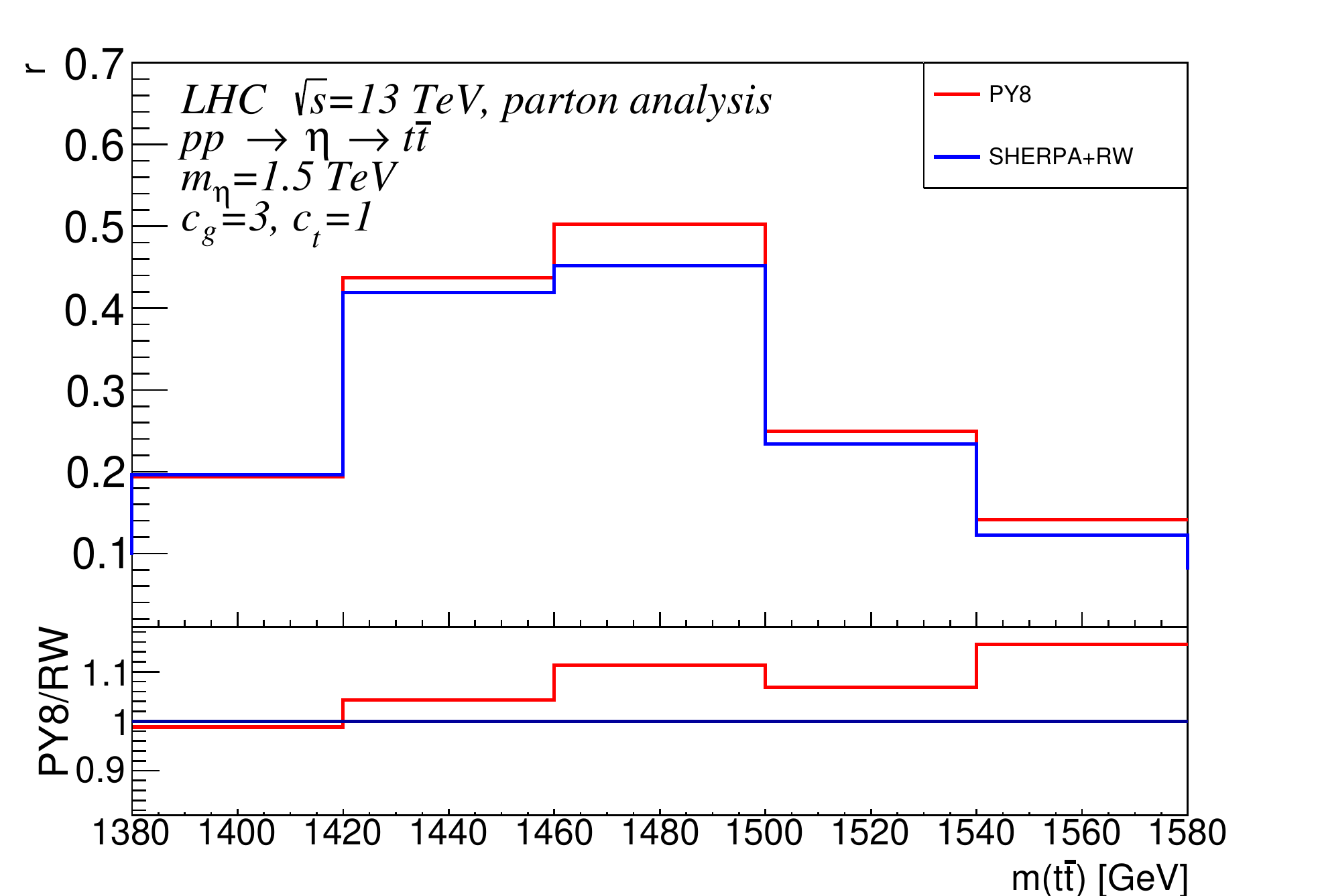}
\caption{Comparison of the predictions for the top-pair invariant mass from the reweighting method (\Sherpa + RW) and
  a dedicated full simulation with \mg5+\textsc{Pythia}8 for a color-singlet pseudo-scalar of mass $m=1.5$ TeV. The \emph{Parton Analysis} was adopted.}
\label{fig:rwgtXfull}
\end{figure} 

%%%%%%%%%%%%%%%%%%%%%%%%%%%%%%%%%%%%%%%%%%%%%%%%%%%%%
\section{Results}
\label{sec:statistics}

Resonant top-quark pair production at the LHC has been analyzed for several of the models mentioned above already.  
Color neutral resonances decaying into $\ttb$ have been studied in several works for a large number of models~\cite{Gaemers:1984sj,Dicus:1994bm,Gori:2016zto,Jung:2015gta,Djouadi:2016ack}, even including interference effects at NLO in QCD~\cite{Bernreuther:2015fts,Hespel:2016qaf,BuarqueFranzosi:2017jrj}.
The case of a color-octet signal has been considered in~\cite{Frederix:2007gi,FileviezPerez:2008ib,Frolov:2016gvu,Hayreter:2017wra}, also considering other production channels, \emph{e.g.} via $b\bar{b}$ initial states, or even double scalar production~\cite{Gerbush:2007fe,Kim:2008bx,Schumann:2011ji}.
Our approach differs from previous studies because we adopt the strategy of directly comparing to data which has been shown
to agree well with the SM prediction, and therefore, can be used to put direct limits on the model parameters,
in the same spirit as~\cite{Butterworth:2016sqg}. 
Indeed, the recent ATLAS measurement of the top-quark pair differential cross section at $\sqrt{s}=13 \TeV$ shows good agreement with various SM Monte Carlo generators~\cite{Aaboud:2017fha}.
However, there are no measurements of the $\ttb$ invariant mass in the boosted regime at this energy yet.
Moreover, the uncertainties are still quite large, since only the 2015 data, corresponding to $3.2\ifb$,
were used, but we expect that an update of the analysis will be available in the near future, with
improved systematics and statistical uncertainty (comparable to the ones presented in this paper)
allowing to derive real exclusion limits. We assume in what follows that data will be well described
by the SM expectation, and take the SM prediction from \Sherpa~ as mock data. 

  The method proposed allows theorists to derive realistic exclusion limits on a variety of NP scenarios
  without a dedicated and expensive experimental analysis. It opens a new path to search for NP,
  with the experiments providing precision measurements of SM processes. With respect to dedicated experimental
  searches, it can serve as check and as an alternative (less-expensive) approach to look for more general
  parametrizations of deviations caused by New Physics. For instance, in the ATLAS and CMS collaborations'
  analyses~\cite{Khachatryan:2015sma,Chatrchyan:2013lca,Chatrchyan:2012cx,Sirunyan:2017uhk,Aaboud:2017hnm}, only
  a leptophobic Z' bosons (present for instance in topcolor scenarios), a Kaluza-Klein excitation of the gluon and heavy states in THDM were searched for. Moreover, interference effects were considered only in Ref.~\cite{Aaboud:2017hnm}.
  With our technique we are able to provide limits for a whole wealth of models.

In order to assess the possibility to observe the signals described above we perform a simple $\chi^2$ analysis
using the bins of the $r$ distribution.  We consider the mass window $m_\phi-200\,{\rm \GeV}< m(t\bar{t})< m_\phi+200\,{\rm \GeV}$ and compute
\begin{equation}
\chi^2_{N}=\sum_{i=1}^{N}\frac{r_i(H)^2}{\sigma_i^2}\,,
\end{equation}
with $N$ the number of bins taken into account, according to the assumed resolution of the measurement.
$r_i(H)$ is the $r(H)$ distribution integrated over bin $i$ and $H$ is the hypothesis (either $S$ or $S+I$). 
$\sigma^2_i$ is the variance on each bin of the distribution.

The variance is derived according to the rules of propagation of uncertainties and is estimated by
\begin{equation}
\sigma^2 = \frac{1}{B_\sigma}\left(1+\frac{H_\sigma^2}{B_\sigma^2}\right) 
+ \epsilon_{\rm SYS}^2\left(1+\frac{H_\sigma^2}{B_\sigma^2}\right) 
+ \epsilon_{\rm TH}^2\frac{(H_\sigma+B_\sigma)^2}{B_\sigma^2} \,.
\end{equation}
We kept the indexes $i$ implicit in the expression. The first term accounts for statistical error, the second for systematic uncertainties of experimental sources, and the third for theoretical uncertainties.
We assume a flat distribution for theory and systematic uncertainty, and that statistical uncertainties are dominated by the background, with a small ratio signal over background.
We take  $\epsilon_{TH}=1\%$ for both $H_\sigma=S_\sigma$ and $H_\sigma=S_\sigma+I_\sigma$, assuming other errors are strongly correlated and will be canceled when taking the ratio distribution. 
%and, for fully uncorrelated uncertainties, is given by 
%\begin{equation}
%(\sigma^{(r)})^2 \sim \frac{\sigma_H^2}{B_\sigma^2} + 
%\frac{\sigma_B^2}{B_\sigma^2}\left(1+\frac{H_\sigma^2}{B_\sigma^2}\right)\,,
%\end{equation}
%where $\sigma_{X}^2$ is the invariance of $X=H_\sigma,\,B_\sigma$. We kept the indexes $i$ implicit in the expression. We assume
%\begin{equation}
%\sigma_X^2 = X  + \epsilon_{\rm X}^2 X^2\,.
%\end{equation}  
%The first term accounts for statistical uncertainty and the second for systematic uncertainties, which we model by a flat distribution. 
%We take $\epsilon_H = \epsilon_{TH}=1\%$ for both $H=S_\sigma$ and $H=S_\sigma+I_\sigma$, which is of small relevance in the analysis. 
The experimental uncertainty is more important and we consider three benchmark estimates for $\epsilon_{\rm SYS}$: 
%
%The variance is derived according to the rules of propagation of uncertainties and, for fully uncorrelated uncertainties, is given by 
%\begin{equation}
%(\sigma^{(r)})^2 \sim \frac{\sigma_H^2}{B_\sigma^2} + 
%\frac{\sigma_B^2}{B_\sigma^2}\left(1+\frac{H_\sigma^2}{B_\sigma^2}\right)\,,
%\end{equation}
%where $\sigma_{X}^2$ is the invariance of $X=H_\sigma,\,B_\sigma$. We kept the indices $i$ implicit in the expression. We assume
%\begin{equation}
%\sigma_X^2 = X  + \epsilon_{\rm X}^2 X^2\,.
%\end{equation}  
%The first term accounts for statistical uncertainty and the second for systematic uncertainties, which we model by a flat distribution. 
%We take $\epsilon_H = \epsilon_{TH}=1\%$ for both $H=S_\sigma$ and $H=S_\sigma+I_\sigma$, which is of small relevance in the analysis. 
%The uncertainty on the SM background is more important. We use $\epsilon_B^2 =  \epsilon_{TH}^2 +  \epsilon_{\rm SYS}^2$ and consider three benchmark error estimates for $\epsilon_{\rm SYS}$: 
\begin{enumerate}
\item In Ref.~\cite{Aaboud:2017hnm} the total systematics on the background were estimated as 10\% and 11\%.
%S+I and 25\%. 
As a pessimistic case we consider $\epsilon_{\rm SYS}=10\%-15\%$. 
\item As an optimistic scenario we vary it to lower values considering a future improved understanding of the uncertainties
  and the reduction in uncertainty associated to normalization. Since we are using a normalized distribution many of the uncertainties estimated in the previous benchmark are strongly correlated and will be canceled out. For this we use
  $\epsilon_{\rm SYS}=5\%-10\%$.
\item As the  most optimistic case we assume experimental uncertainties can be drastically reduced to the level of theoretical, which according to Ref.~\cite{Czakon:2017wor} results in $\epsilon_{\rm SYS}=1\%-2\%$. 
\end{enumerate}

We consider $N=1$ for a \emph{bad} resolution case, assuming the experiment can resolve only the full window of 400 GeV in $m(t\bar{t})$, and $N=10$ assuming a mass resolution in $m(t\bar{t})$ of $40$ GeV.

% including both theoretical (scale dependence, reweighting, etc.) and experimental sources. 
%The variance of $d\sigma_H/dm$ is given by $\sigma^2
%, $S$ the signal prediction for each bin, $B$ the background prediction and $\sigma$ the associated uncertainty, which we estimate as
%
%\begin{equation}
%\sigma^2_i=B+\epsilon_{\rm SYS}^2 B^2\,.
%\end{equation}
%%
%The first term accounts for statistical error and the second for systematic uncertainties, including both theoretical (scale dependence, reweighting, etc.) and experimental sources. 
We consider $\chi^2\geq 2$ as a criterion for exclusion, which corresponds roughly to an exclusion at 95\% of confidence level.
%We did not compare the distribution with real data \com{because it is not available for LHC RunII}, but only with the expected SM prediction. 

This simple analysis is intended to be a first approximation to a full statistical data analysis that will be carried
out eventually. In particular we assume the same uncertainty for every bin without correlation between them, and we assume only two
cases of resolution independent of the bin. In the following we discuss some benchmark scenarios and the respective results.

%%%%%%%%%%%%%%%%%%
\subsection{Pseudo-scalar color octet}

The first scenario we consider is when the resonance $\phi$ represents a pseudo-scalar color octet ($\widetilde{\eta}$) with total width dominated by the decays to pairs of tops and gluons
\begin{equation}
\Gamma_{\rm TOT}=\Gamma_{tt}+\Gamma_{gg}\,.
\label{eq:decaywidth}
\end{equation}
In Fig.~\ref{fig:Octet500} we show the resulting $r$ distribution assuming a color octet resonance with mass  $m_{\tilde \eta}=500$ GeV and the parameters $c_t=1$, $c_g=1$ (left) and $c_g=-1$ (right) at parton level, \emph{i.e.} using the \emph{Parton Analysis}
described in \sec{sec:analysis}. We show both the full line-shape, which comprises signal and interference with QCD background (S+I), and the pure signal (S) for comparison. The importance of
taking into account interference effects can clearly be noticed.
\begin{figure}[H]
\includegraphics[width=0.49\textwidth]{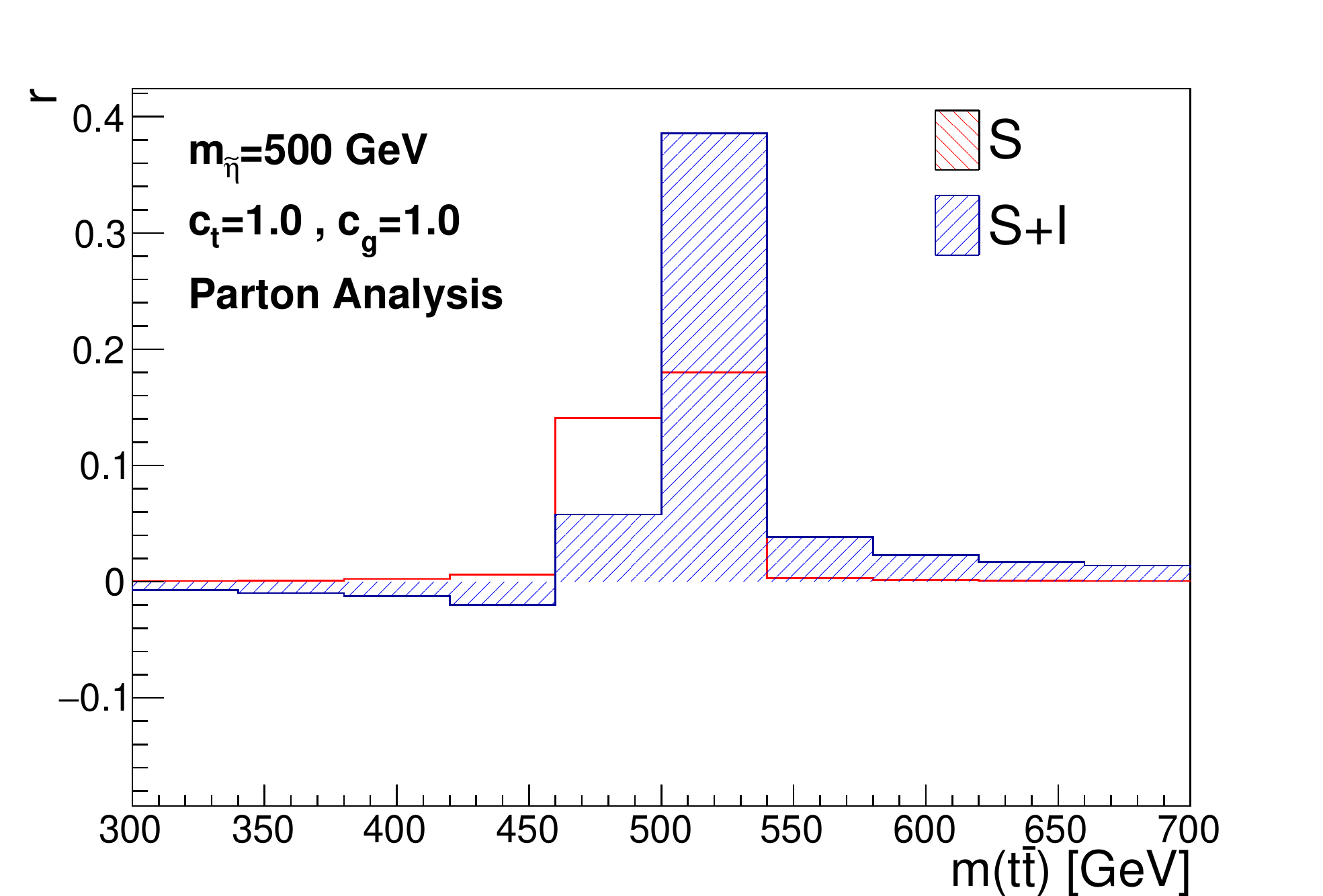}
\includegraphics[width=0.49\textwidth]{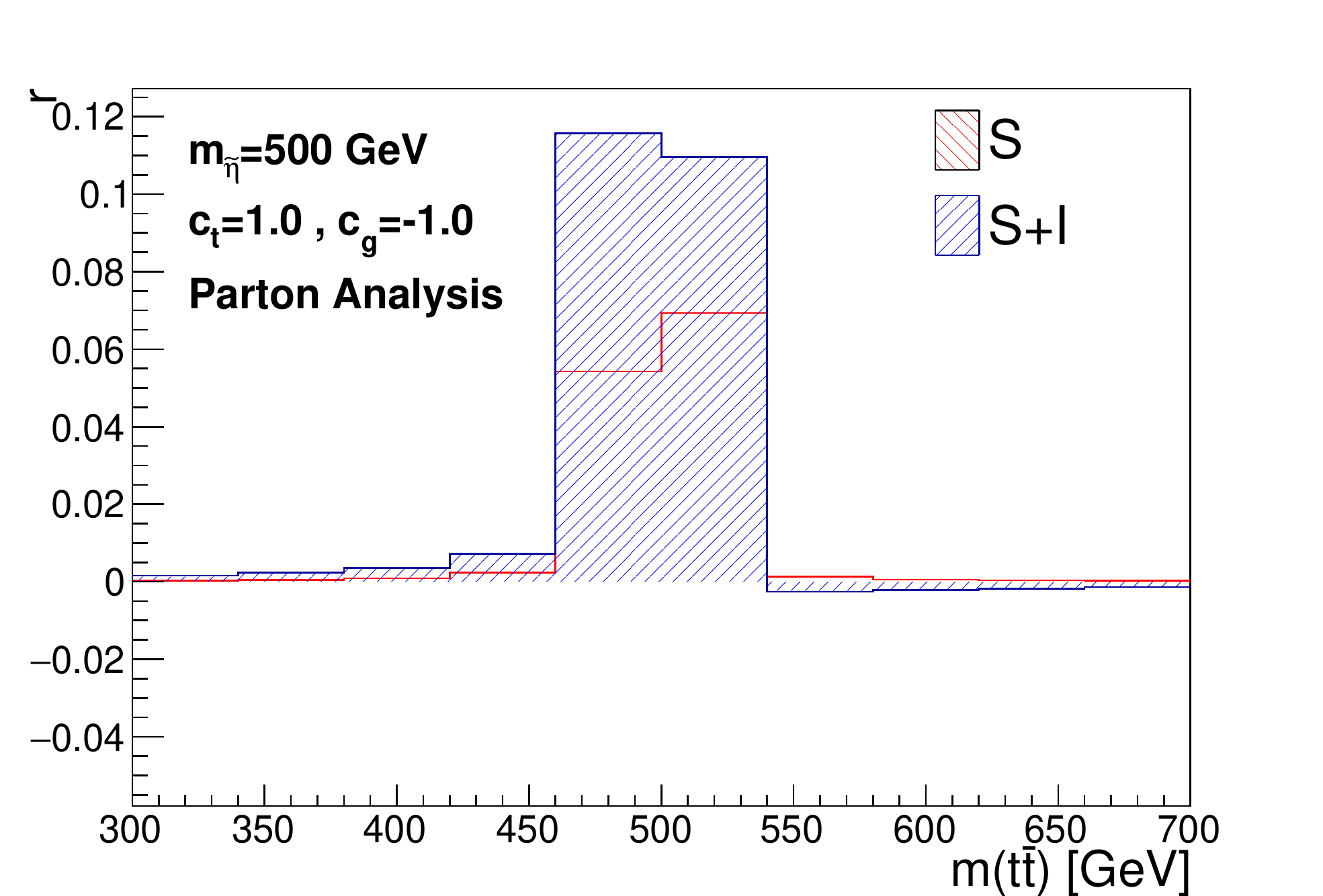} 
\caption{Normalized top-pair mass distributions, $r\equiv\frac{d\sigma/dm}{d\sigma_{\rm SM}/dm}$ for a pseudo-scalar color octet resonance with $m_{\tilde\eta}=500\,{\rm \GeV}$, $c_t=1$ and $c_g=1$ ($c_g=-1$) on the left (right) using the \emph{Parton} analysis. Signal plus interference (S+I) is in blue and pure signal (S) in red.}
\label{fig:Octet500}
\end{figure}

Similarly, in Fig.~\ref{fig:Octet1700}, we present the effect of a resonance with mass $m_{\tilde \eta}=1700\,{\rm GeV}$ and couplings $c_t=1$, $c_g=1$ (left) and $c_g=-1$ (right), reconstructed using the \emph{Boosted Analysis}. 
The excess reaches more than 10\%, which indicates that even a pessimistic estimate of the uncertainties is sufficient to exclude the existence of this state for values of $c_g$ of order 1. We thus use the most pessimistic value for the systematic error, $\epsilon_{\rm SYS}=10\%-15\%$.
\begin{figure}[H]
\includegraphics[width=0.49\textwidth]{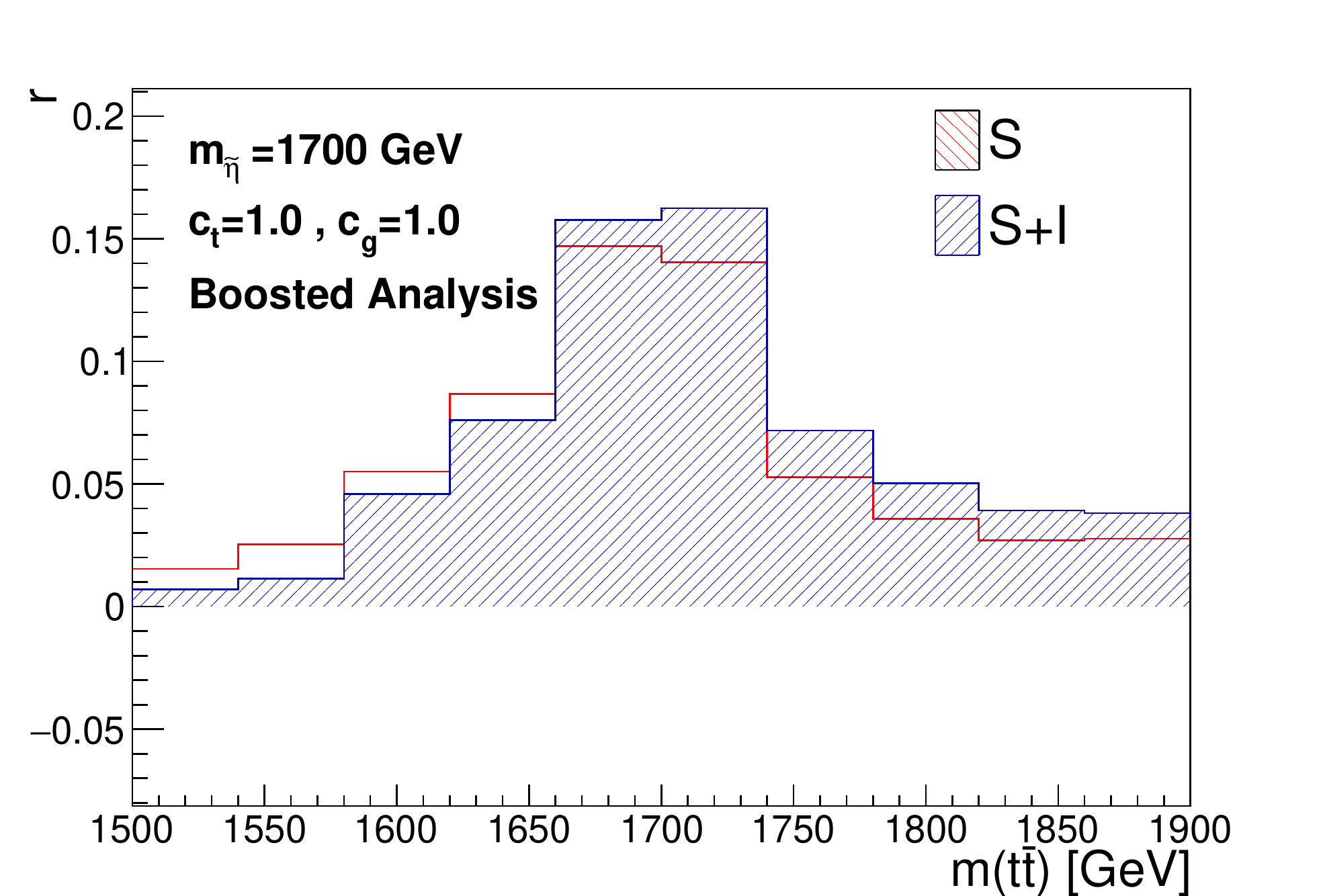}
\includegraphics[width=0.49\textwidth]{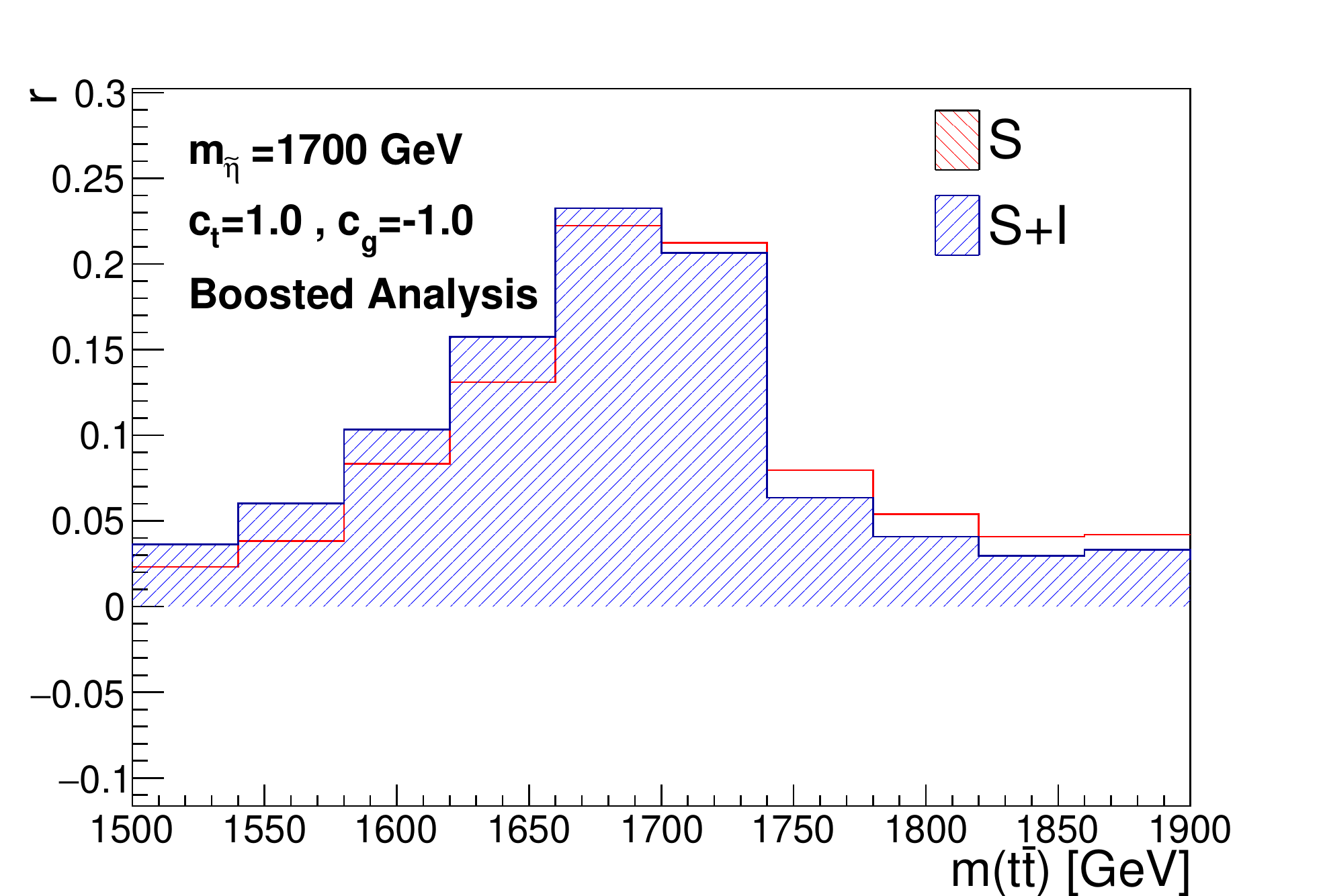} 
\caption{Normalized top-pair mass distributions $r$ reconstructed with the \emph{Boosted} analysis for a pseudo-scalar color octet resonance with $m_{\tilde \eta}=1700\,{\rm \GeV}$, $c_t=1$ and $c_g=1$ ($c_g=-1$) on the left (right). The color scheme is the same as in Fig.~\ref{fig:Octet500}.}
\label{fig:Octet1700}
\end{figure}

In Fig.~\ref{fig:Octetlimit} the corresponding exclusion limits are shown, assuming a fixed value of $c_{t}=1$. The bands correspond to a systematic uncertainty on the measurement running from $10\%$ to $15\%$.
The limits are evaluated considering the interference effect (dashed lines) or neglecting it (continuous lines). The interference has a significant effect in the low mass region ($m_{\tilde{\eta}} < 1.3\,{\rm TeV}$).
The excluded region corresponds to larger values of $|c_g|$. We show the exclusion for integrated luminosities of $L=20\ifb$ (blue line) and $L=100\ifb$ (black). In the \emph{left-panel} we use $10$ bins of $40\,{\rm GeV}$ width in the invariant-mass distribution to compute $\chi^2_{10}=2$ while on the \emph{right-panel} we use only a single 400 GeV bin centered around the resonance mass, $\chi^2_{1}=2$. 
The comparison between the left and right panel shows the importance of a good resolution and for a line-shape analysis.
\begin{figure}[H]
\includegraphics[width=0.49\textwidth]{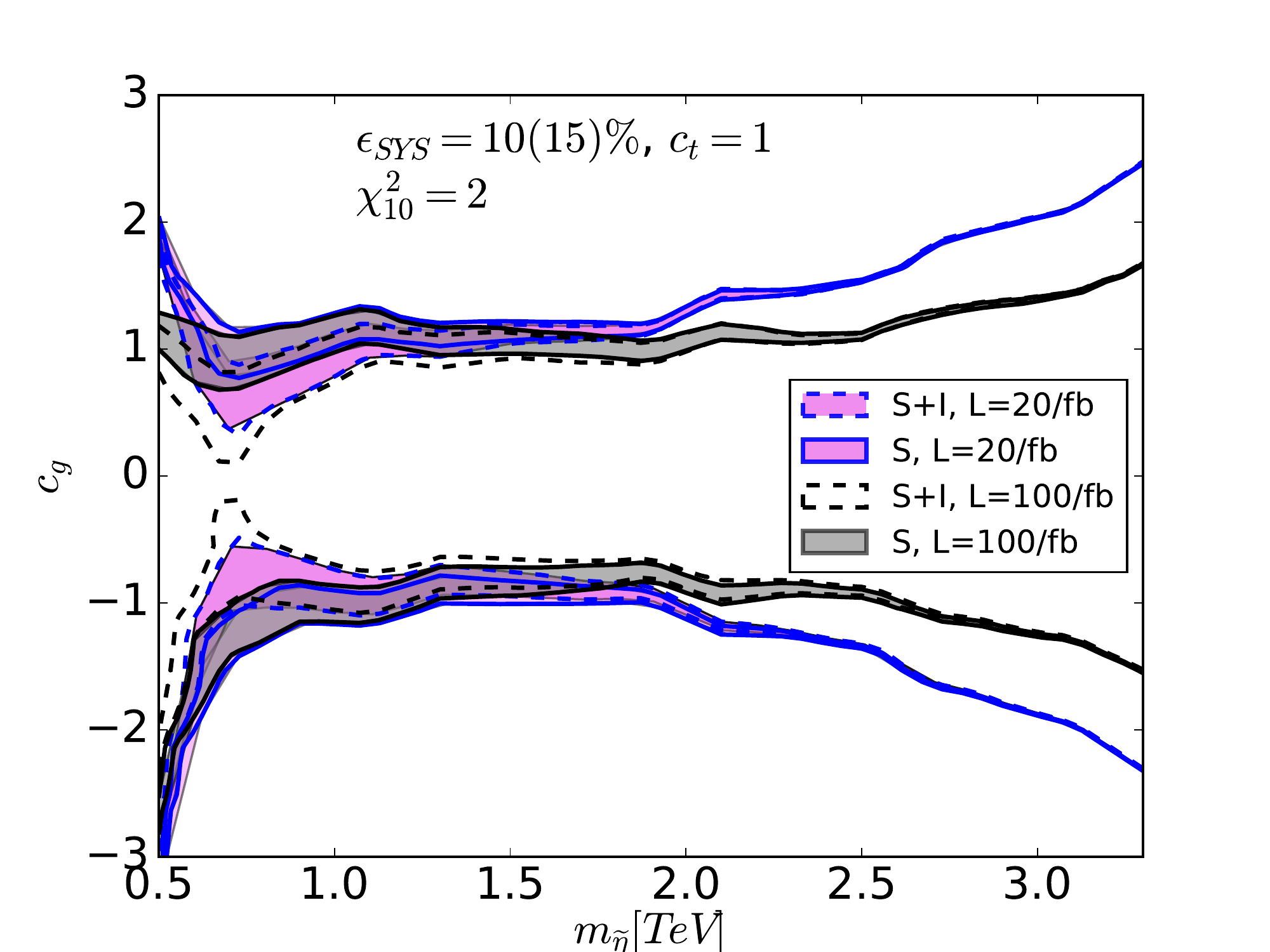}
\includegraphics[width=0.49\textwidth]{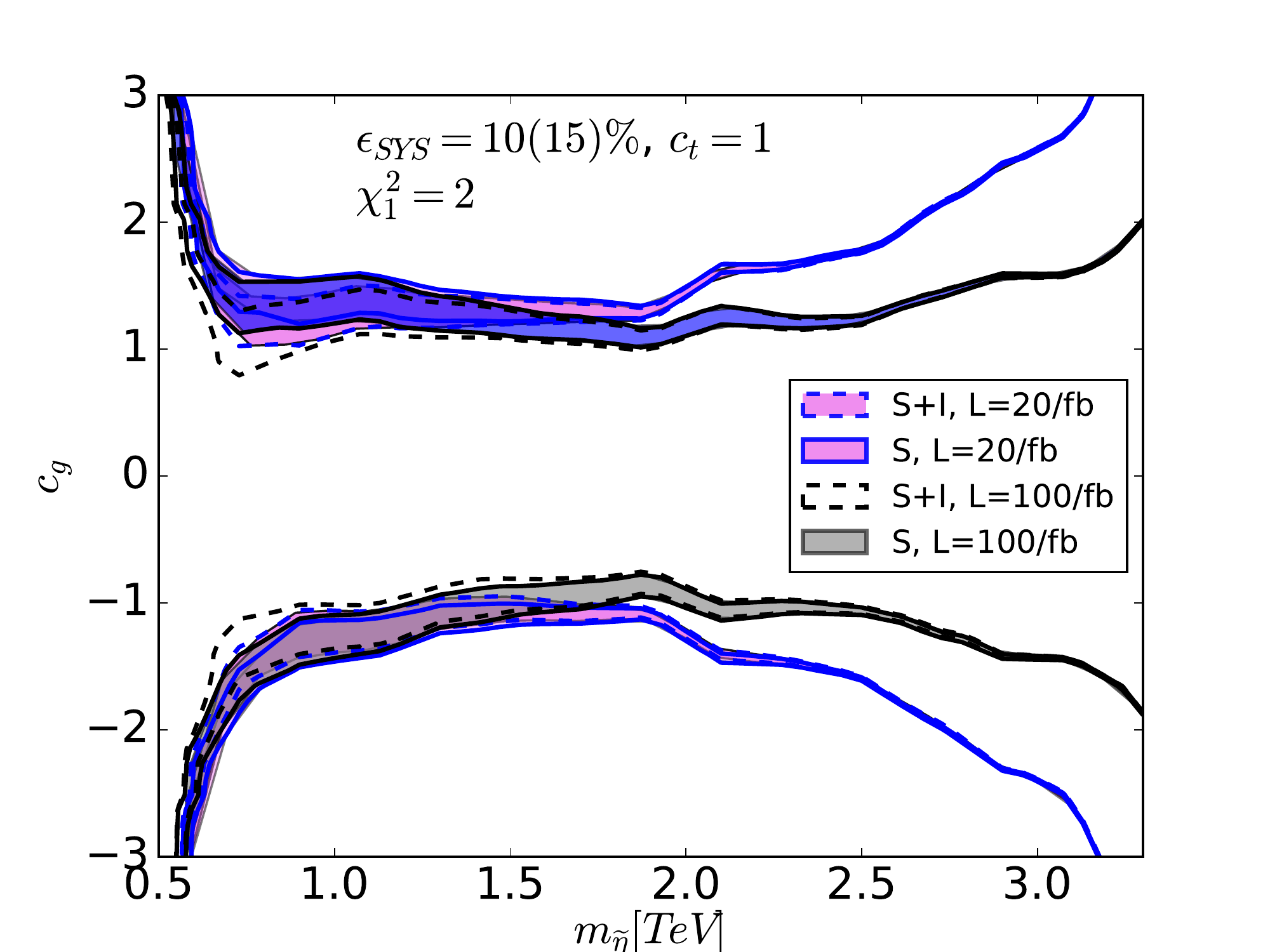} 
\caption{Exclusion limits ($\chi^2=2$) in $(m_{\tilde \eta},c_g)$ parameter space for a pseudo-scalar color octet assuming $c_t=1$. The band represents the different assumptions for the systematic uncertainty, varying from 10\% to 15\%. Integrated luminosities are $L=20\ifb$ (blue line) and $L=100\ifb$ (black), as well as considering interference (dashed line) and neglecting it (solid line).}
\label{fig:Octetlimit}
\end{figure}

We expect striking signatures in other channels, but little has been studied. For instance, in the analysis of $\gamma$+jets in Ref.~\cite{Aaboud:2017nak} a color octet has not been considered.

%%%%%%%%%%%%%%%%%%%%%%%%%%%%%%%%%%%%%%%%
\subsection{Pseudo-scalar singlet}

For the benchmark scenario of a pseudo-scalar color singlet we again assume the resonance' width is dominated by the top and
gluon decays, as in \eq{eq:decaywidth}.

We show in Fig.~\ref{fig:Boostedm1500} the distribution of the normalized $m(t\bar{t})$ distribution $r$ assuming $m_\eta=1500\,{\rm GeV}$. In the left-hand (right-hand) panel we consider $c_g=1$ ($c_g=-1$). The line-shapes of this scenario are highly non-trivial,
they strongly depend on the mass and couplings, and can feature pure dips, pure peaks and intermediate peak-dip or dip-peak structures. A sample of different line-shapes is shown in \app{app:lineshapes}.

\begin{figure}[H]
\includegraphics[width=0.49\textwidth]{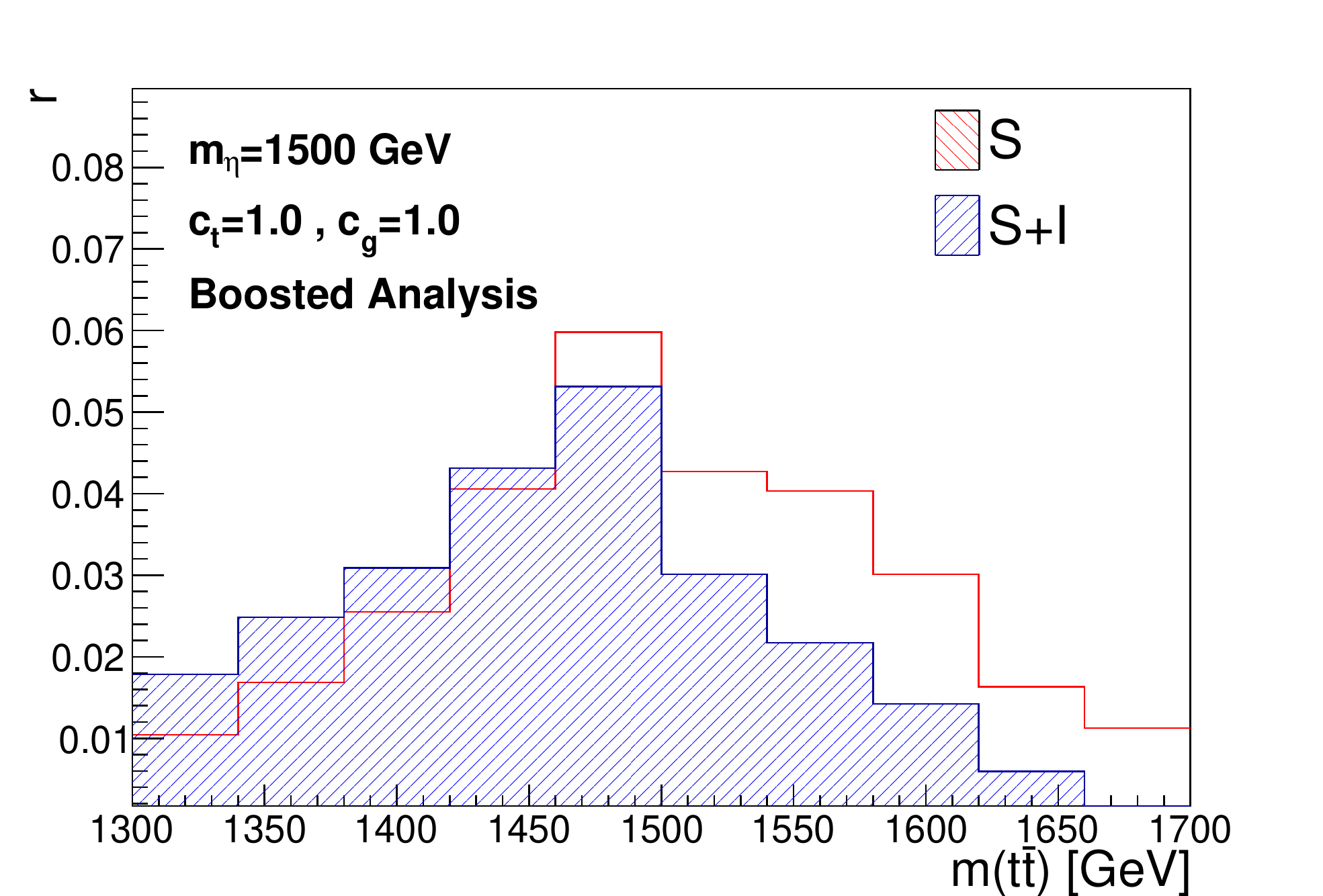}
\includegraphics[width=0.49\textwidth]{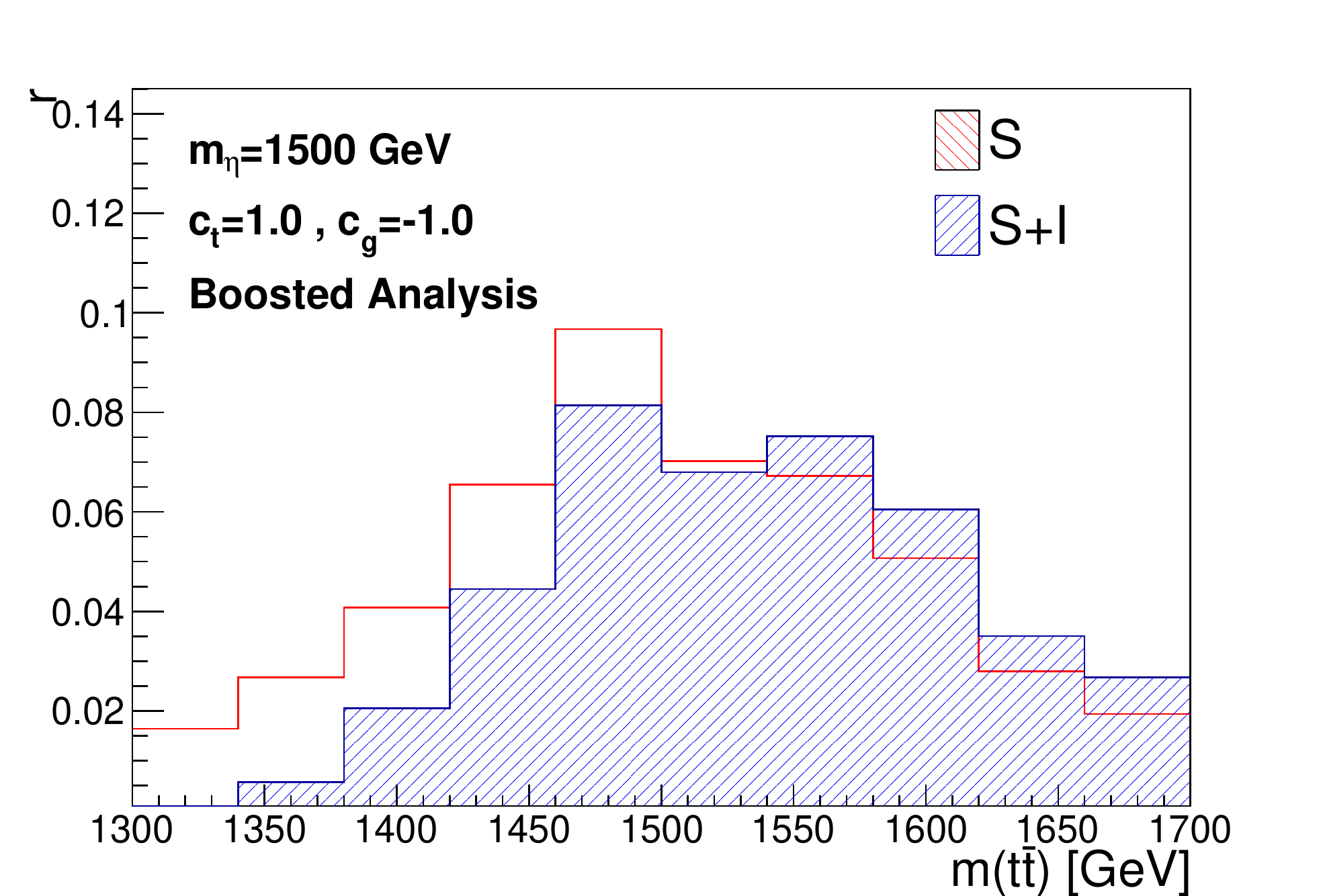} 
\caption{Normalized top-pair mass distributions $r$ reconstructed with the \emph{Boosted} analysis for a pseudo-scalar color singlet resonance with $m_\eta=1500\,{\rm \GeV}$, $c_t=1$ and $c_g=1$ ($c_g=-1$) on the left (right). The color-scheme is the same as in Fig.~\ref{fig:Octet500}.
}
\label{fig:Boostedm1500}
\end{figure}

%If $c_g=0$, \emph{i.e.} only the top loop contributes to the production, and the signal is very small to be observed, as can be seen in \figs{}{}. The difference is always less than the sytstematic error of 5\% we assumed. A more efficient analysis, other top decay channels will help in improving this situation. In \fig{} we show the significance ($\sqrt{\chi^2}$) for this scenario. The realistic analysis have much lower significance compared to the parton level case.
%For larger mass statistics becomes important. We also show the importance of having a good resolution. Using bin =40 GeV and doing a chi2 we get the left panel. Doing just a usual bump significance with one large bin we get the right panel (this could be improved by choosing better the window). 

%% Exclusions 
In Fig.~\ref{fig:Exclct1} we show the exclusion limits in the $(m_\eta,c_g)$ parameter space plane  for $c_t=1$. The band represents the different assumptions for the systematic uncertainty, 5\% and 10\%. 
The effect of interference is important for low masses $m_\eta\lesssim 1.2\,{\rm \TeV}$, where also systematics dominate and have a huge impact on the exclusion power. The use of the full line-shape in the statistical analysis improves the exclusion power mostly for low masses where more distinct line-shapes are present. For masses above $m_\eta\gtrsim 2\,{\rm \TeV}$, higher luminosities than $L=100\ifb$ are needed.

In Fig.~\ref{fig:Exclm1500} we show the corresponding exclusion limits in the $(c_t,c_g)$ plane for a fixed mass $m_\eta=1.5\,{\rm \TeV}$. 
The effect of interference is important for large top couplings, $c_t\gtrsim 1.2$, which is directly related to the size of the width. The use of full line-shape gives a mild improvement in the exclusion power.

\begin{figure}[H]
\includegraphics[width=0.49\textwidth]{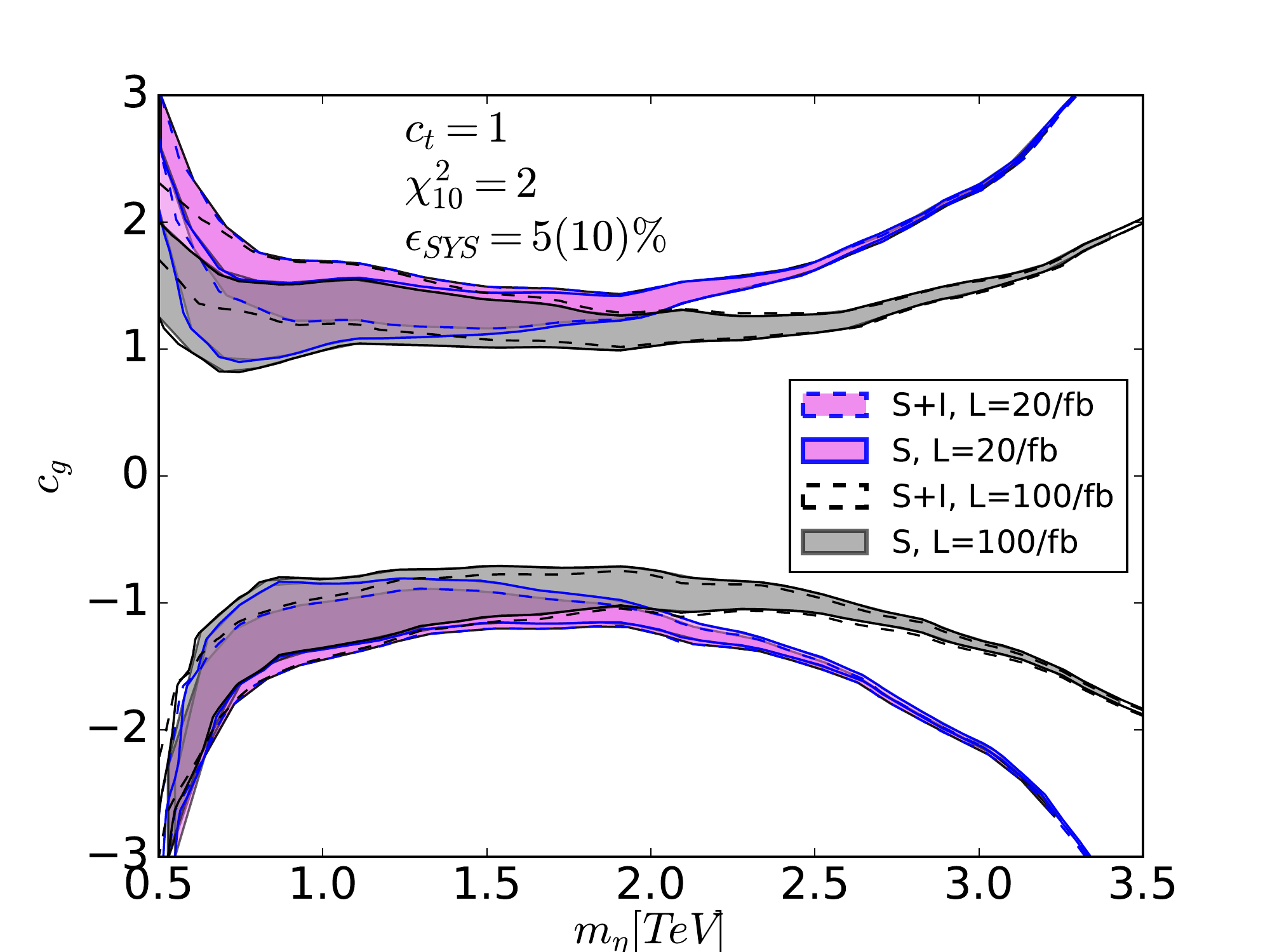}  
\includegraphics[width=0.49\textwidth]{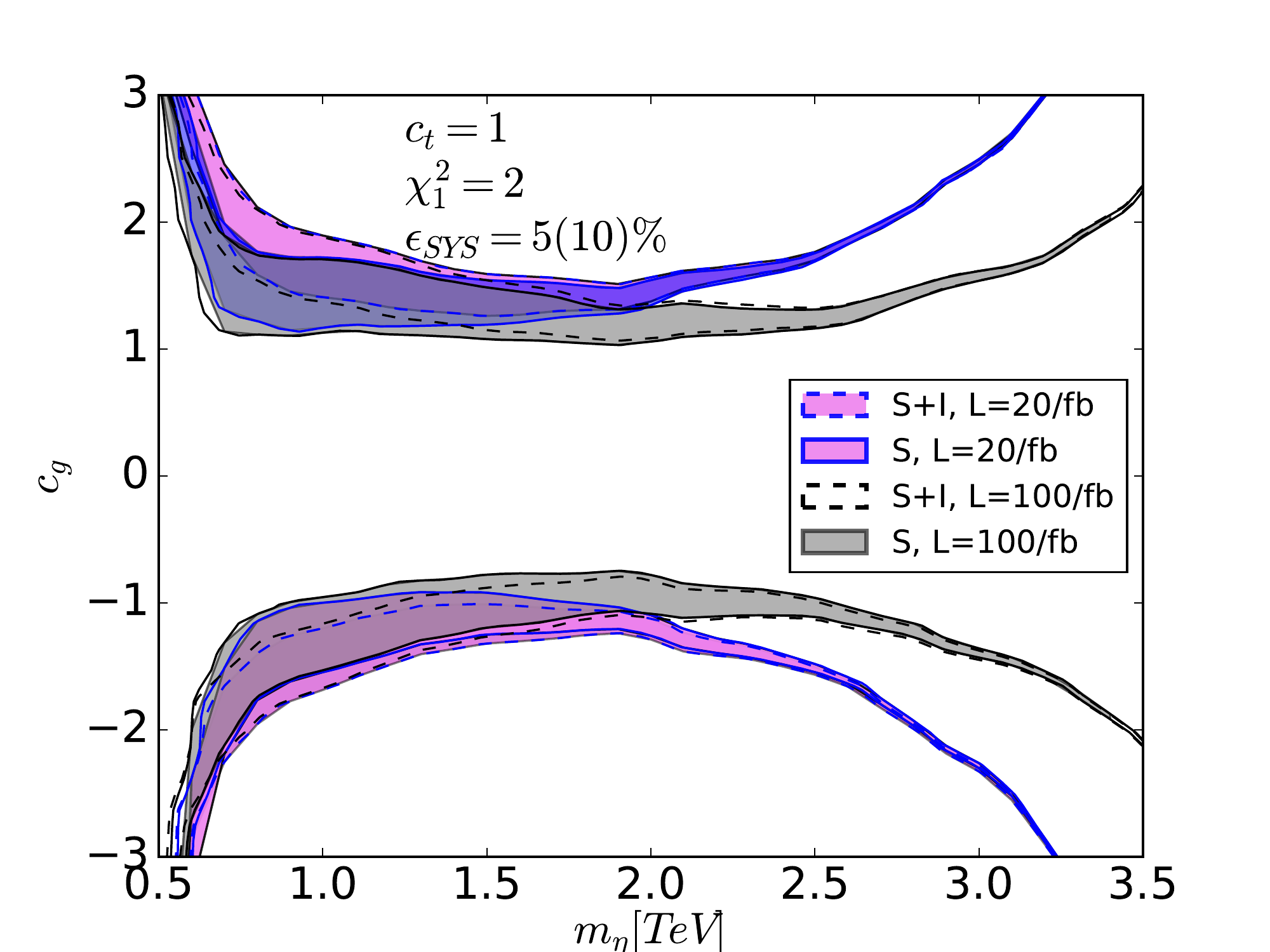} 
\caption{Exclusion limits ($\chi^2=2$) in $(m_\eta,c_g)$ parameter space and $c_t=1$ for a pseudo-scalar color singlet. The band represents the different assumptions for the systematic uncertainty, varying from 5\% to 10\%. The color and style scheme for the lines are the same as in Fig.~\ref{fig:Octetlimit}. }
\label{fig:Exclct1}
\end{figure}

\begin{figure}[H]
\includegraphics[width=0.49\textwidth]{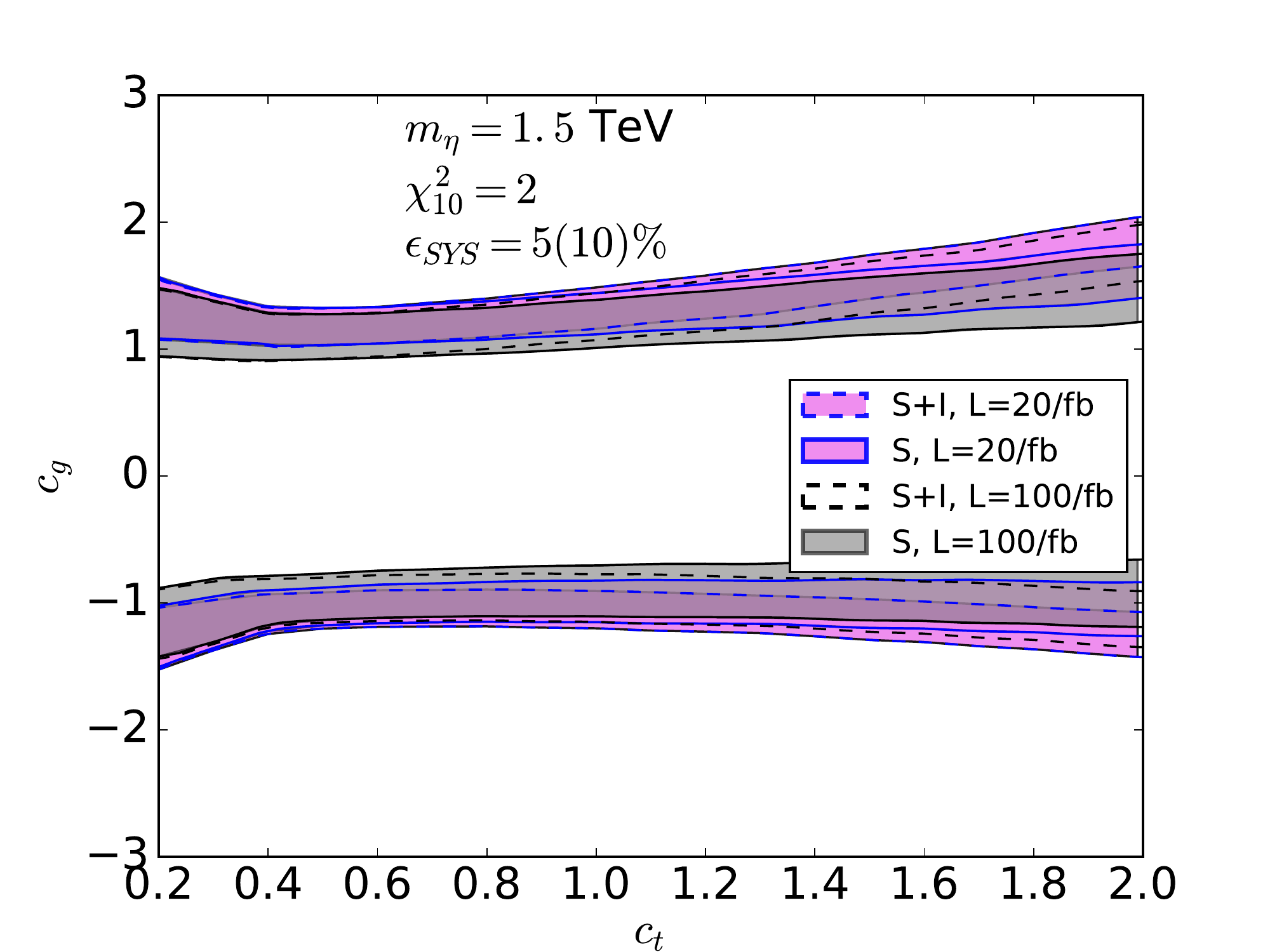}  
\includegraphics[width=0.49\textwidth]{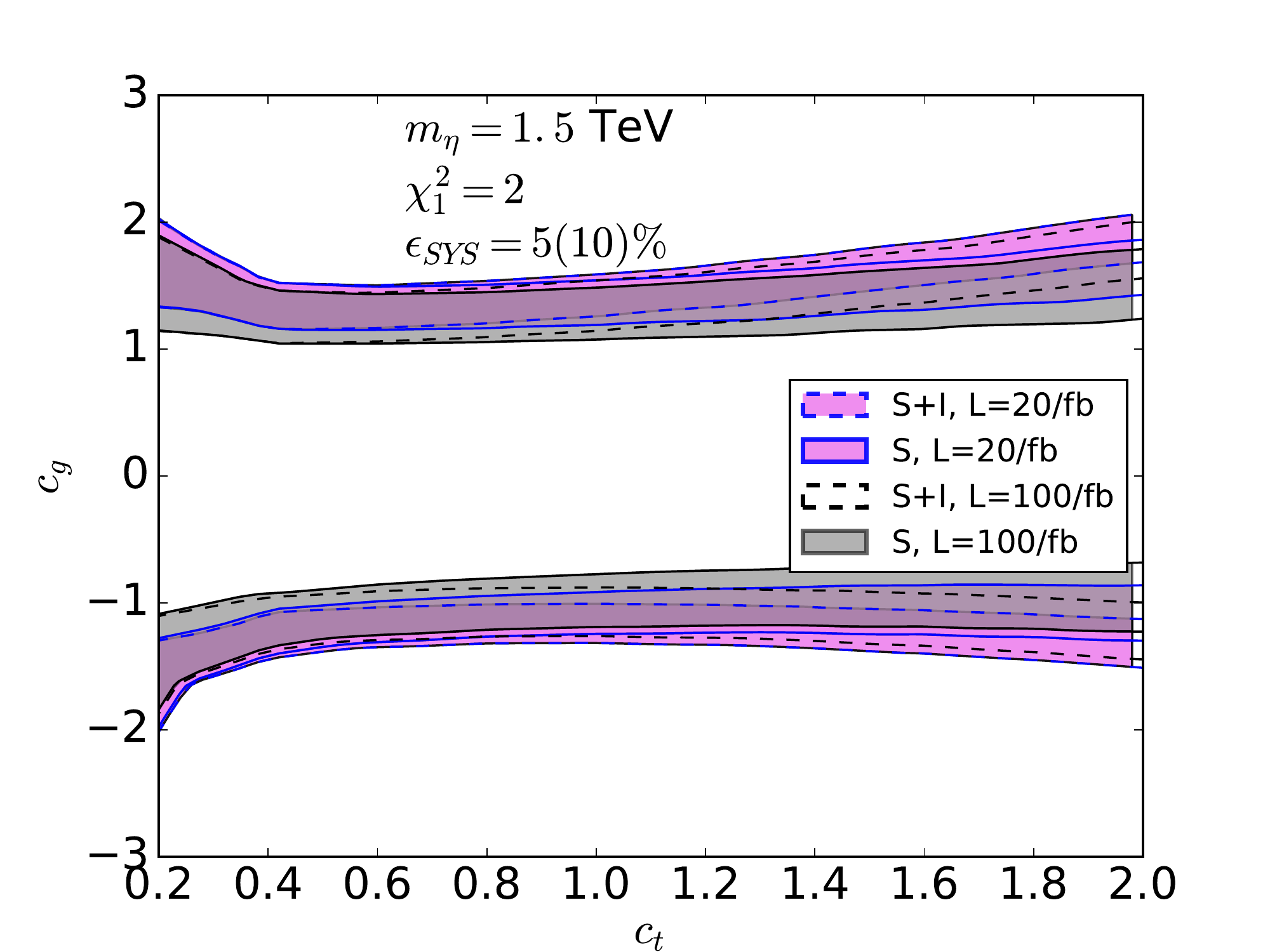} 
\caption{Equivalent to Fig.~\ref{fig:Exclct1} for the $(c_t,c_g)$ plane for a fixed mass of $m_\eta=1.5\,{\rm \TeV}$ }
\label{fig:Exclm1500}
\end{figure}

For very low masses the \emph{Resolved} analysis can be slightly more powerful than the \emph{Boosted}.
In Fig.~\ref{fig:Resolved} on the left we show an example of a line-shape and on the right the exclusion limit
provided by the \emph{Resolved} analysis.
Compared to \fig{fig:Exclct1} it can be noticed that the low mass region  $m_\eta\lesssim 600\GeV$ can be better covered by the \emph{Resolved} selection. We note as well that the case of negative $c_g$ is less excluded due to the fact that larger cancellations between top-quark loop and effective vertex happens for these masses.

\begin{figure}[H]
\includegraphics[width=0.49\textwidth]{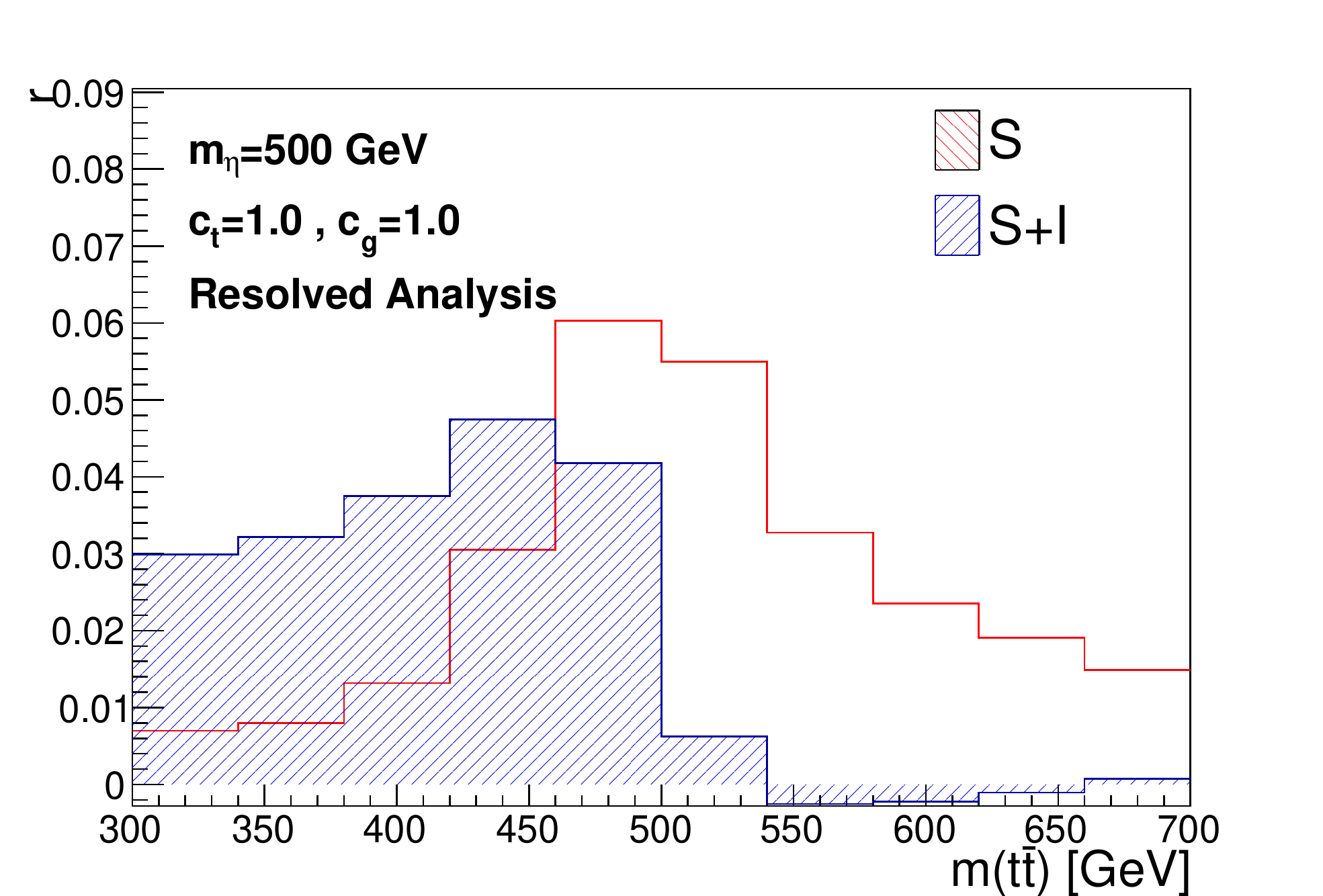}  
\includegraphics[width=0.49\textwidth]{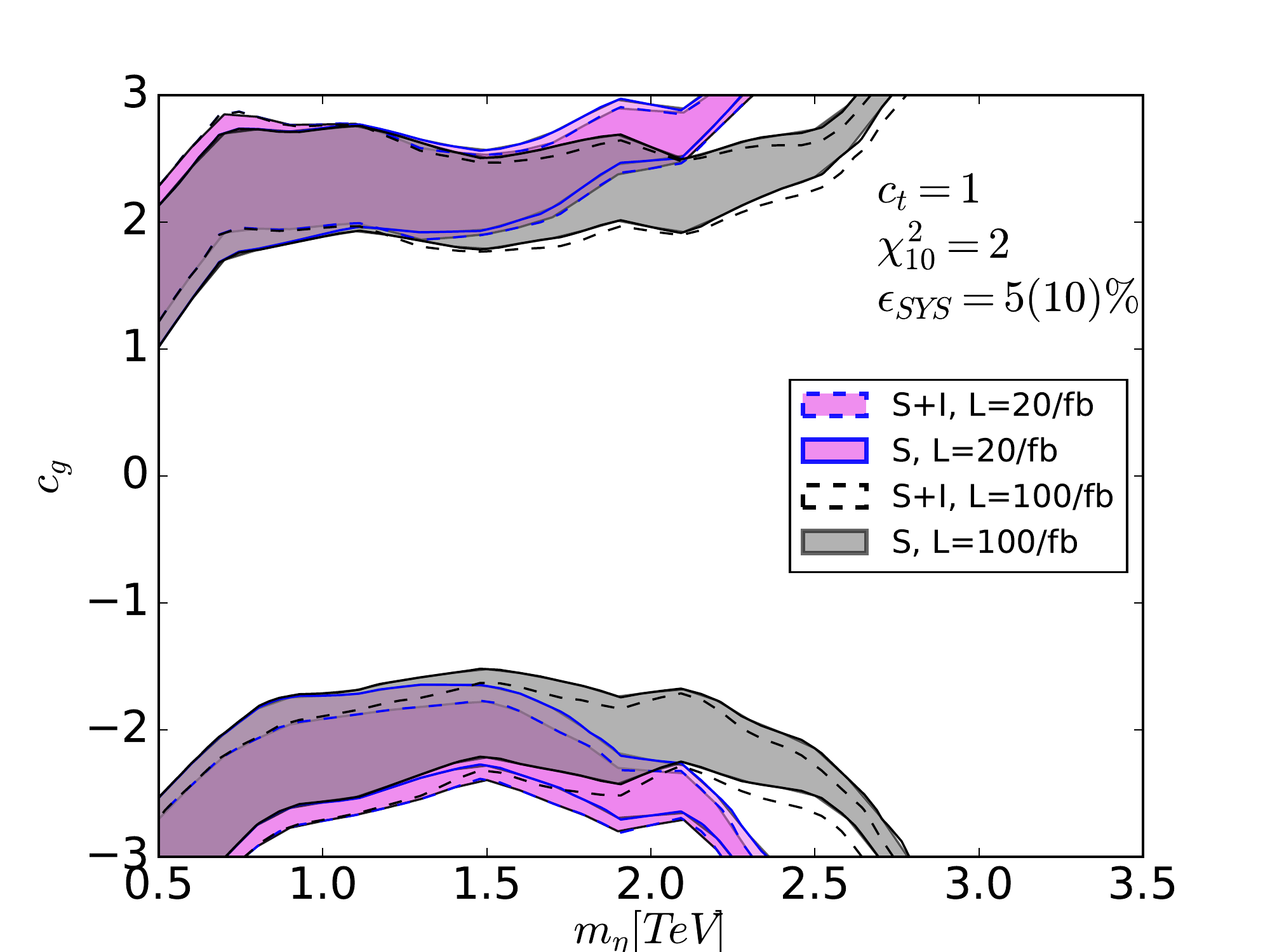}  
\caption{\emph{Left:} Normalized top pair mass distributions $r$ for $c_g=c_t=1$ and $m_\eta=500\,{\rm \GeV}$. \emph{Right:} Exclusion limit ($\chi^2=2$) in $(m_\eta,c_g)$ parameter space for $c_t=1$. The color scheme is the same as in Fig.~\ref{fig:Exclct1}. In both panels the \emph{Resolved} analysis has been employed.}
\label{fig:Resolved}
\end{figure}

 Diphoton and dijet searches might be relevant in extreme regions of parameter space, \emph{i.e.}
  for very small $c_t\sim 0.2$, and large masses, due to the dependence of the $\gamma\gamma$ and $gg$ partial widths
  on $m^3$ as opposed to the linear dependence of the $t\bar{t}$ decay width. 
  In \fig{fig:dijet-diphoton} we show the 95\%CL excluded region derived from the limits provided by the ATLAS
  collaboration in the dijet search~\cite{Aaboud:2017yvp}. We used the case $\sigma_G/m_G=0$ and assumed
  an acceptance of 50\%. In the same figure we show the 95\%CL excluded region in the diphoton channel using
  the exclusion limits by the ATLAS analysis in Ref.~\cite{Aaboud:2017yyg}. We used the case $\Gamma_X/M_X=6\%$
  and the spin-0 selection. To derive cross sections we used the N$^3$LO result for Higgs production cross section
  $\sigma_h$~\cite{Anastasiou:2016hlm} and rescale by the LO decay width,
\begin{equation}
\sigma_\eta = \sigma_h\frac{\Gamma_{\eta\to gg}}{ \Gamma_{h\to gg}}= 
\sigma_h\frac{\left|  c_t^\eta A^{A}_{1/2}\left(\frac{m_\eta^2}{4m_t^2}\right)+ c_g^\eta \right|^2}{\left|A^{S}_{1/2}\left(\frac{m_\eta^2}{4m_t^2}\right)\right|^2}.
\end{equation}
$\Gamma_{\eta\to gg} $ is given in \eq{eq:gamma-eta-gg} and the form factors in eqs.~(\ref{eq:AA}--\ref{eq:ftau}).
The shaded area in the figure represents the region where $\sigma_\eta\times$BR is larger than the excluded line in the respective references, and BR is the corresponding branching ratios. 
%In the shaded area in the figure $\sigma_\eta$ is larger than the excluded cross section provided by the ATLAS experiment.
We can notice that these channels get competitive in sensitivity to $t\bar{t}$ analysis at low $c_t$ and large mass, but only if $c_\gamma$ is particularly large.
In particular, even for $c_t=1$, for $m>3 \TeV$ the dijet search seems to be more sensitive to New Physics.

\begin{figure}[H]
\includegraphics[width=0.49\textwidth]{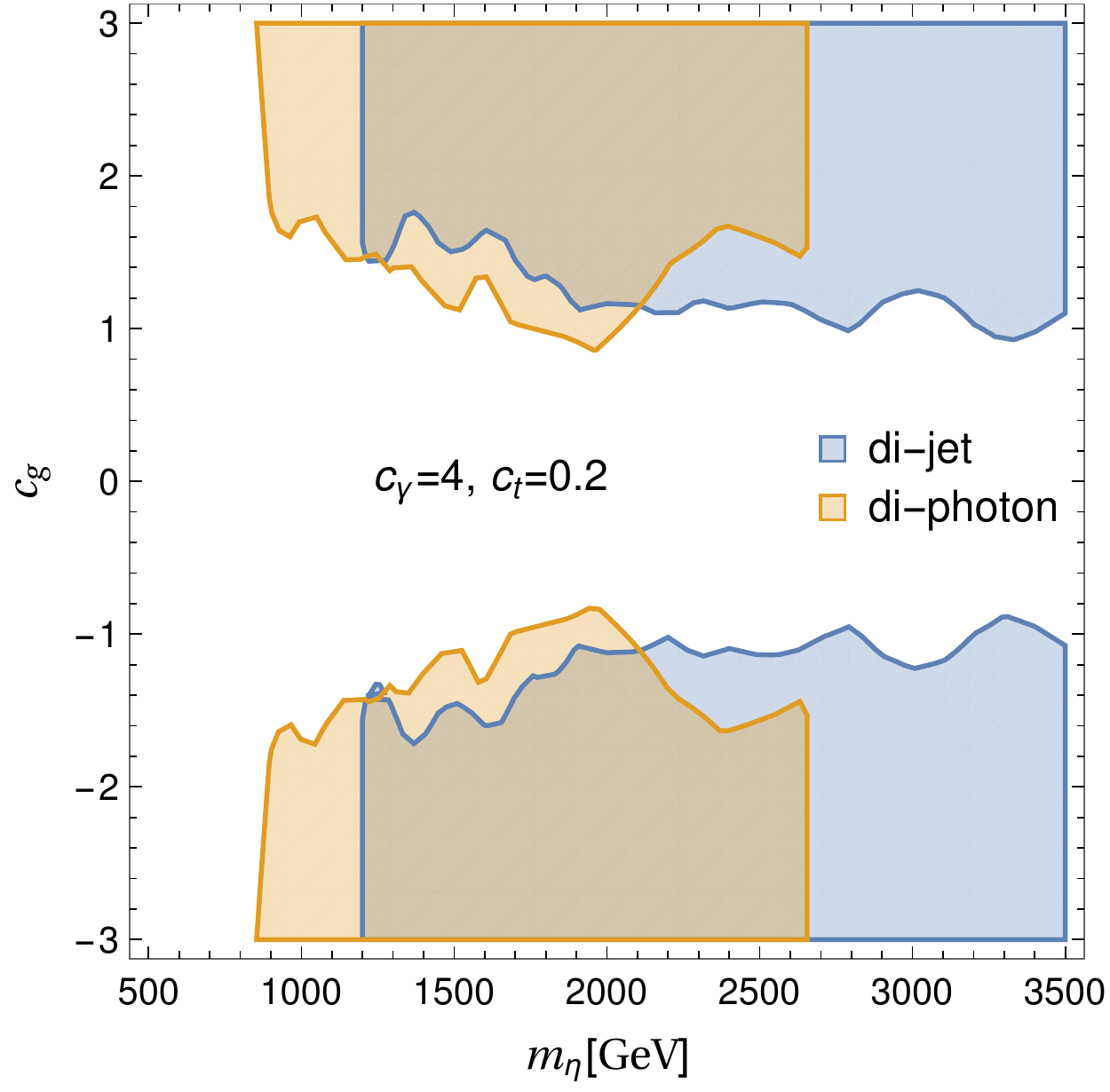}  
\includegraphics[width=0.49\textwidth]{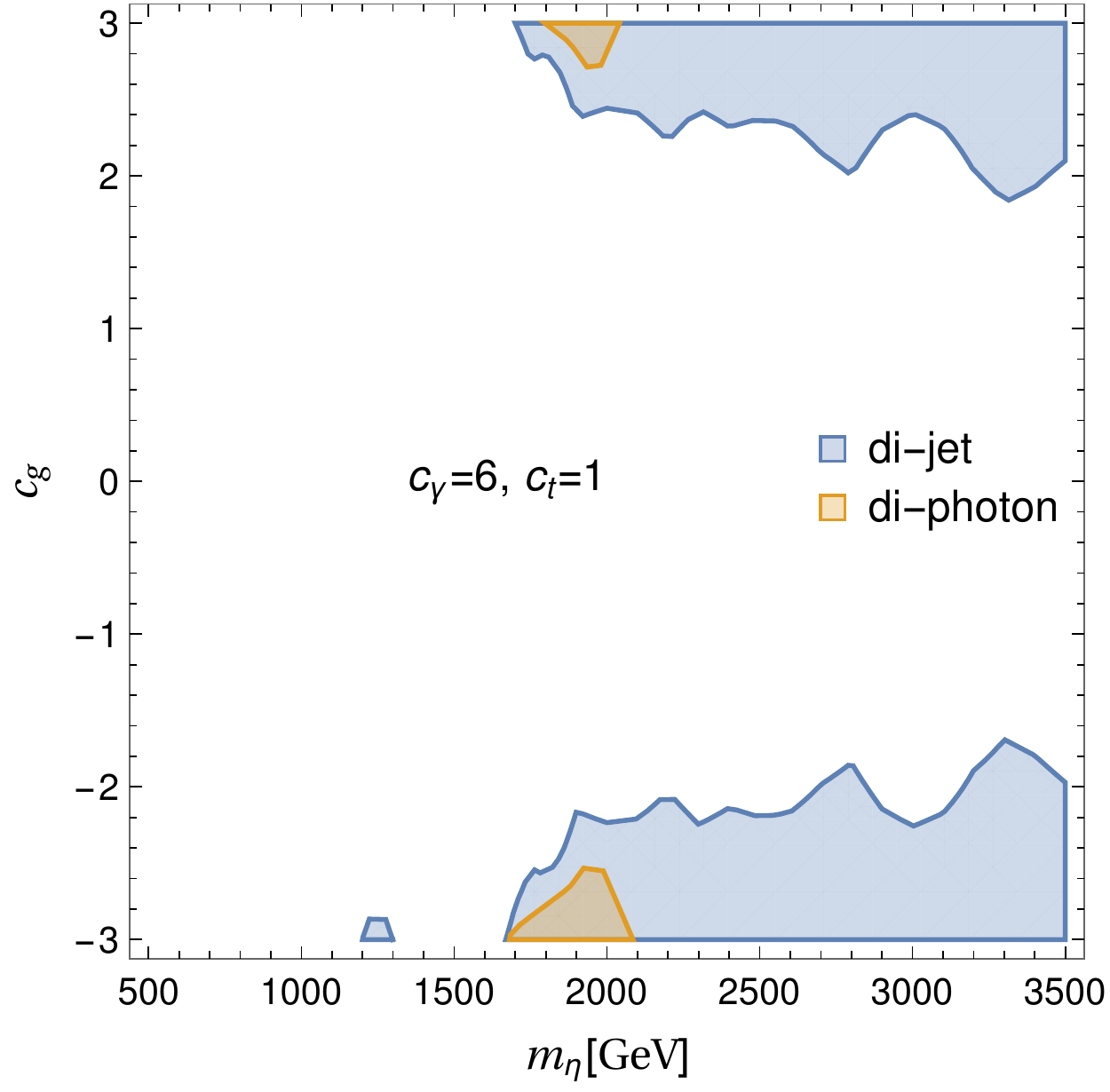}  
\caption{95\%CL excluded region in parameter space in diphoton~\cite{Aaboud:2017yyg} and dijet searches~\cite{Aaboud:2017yvp}. On the \emph{left panel} $c_\gamma=4$ and $c_t=0.2$. On the \emph{right}, $c_\gamma=6$ and $c_t=1$.   }
\label{fig:dijet-diphoton}
\end{figure}

%%%%%%%%%%%%%%%%%%%%%%
\subsubsection*{Interpretation for Composite Higgs models with Top Partial Compositeness}

As an ultra-violet realization of the pseudo-scalar scenario we consider the composite models M3, M8 and M9 of Ref.~\cite{Belyaev:2016ftv}. These models are  constituted by two additional confining fermions, $\psi$ and $\chi$, which form several composite states among which a top partner that can generate a mass to the top quarks through the partial-compositeness mechanism. In addition, they present two iso-singlet pseudo-scalar mass eigenstates $a$ and $\eta'$.
In general, the observation of such pseudo-scalar state decaying into top quarks can shed light on the mechanism of fermion mass generation~\cite{Alanne:2016rpe}.
These models present extra parameters which determine the couplings, given by a pair of integers $(n_\psi,n_\xi)$ and the relation between the mixing angle $\alpha$ and the ratio of scales and U(1) charges, $\zeta$. 
We do not enter a discussion of the details of these the models and their parameters here but invite the reader to consult Ref.~\cite{Belyaev:2016ftv}.
We choose $\alpha=\zeta$ and the values of $(n_\psi,n_\xi)$ which provide the largest couplings to the tops, $(n_\psi,n_\xi)=(2,0),\,(-4,2)$ and $(4,2)$. 
We neglect contributions to the resonance width from the decays into $Z$, $W$ and $\gamma$, which are sub-dominant. 
The relevant couplings are summarized in Tab.~\ref{tab:models}.
%For more detail and explanation about these parameters we refer the reader to the publication.

%\begin{table}
%\caption{
%Summary of the composite models here considered. $c_t$ and $c_g$ are given in units of $v/F_\pi$. $c_t$ is shown for the three cases considered, $(n_\psi,n_\xi)=(2,0)/(-4,2)/(4,2)$. }
%\label{tab:models}
%\begin{tabular}{ l c c c }
% model 		& state 		& $\quad c_t^{\eta,\sigma}[v/F_\pi]\quad$ & $c_g^{\eta,\sigma}[v/F_\pi]$ \\
% \hline\hline
%M3			& a			& .934/1.09/-2.65		& 5.44	\\
%%			& $\eta'$	& .457/1.99/.162			& -7.54	\\
%M8			& a			& .926 /1.54/-2.16		& $1.54$	\\
%%			& $\eta'$	& .193/1.19/.416			& -4.0	\\
%M9			& a			& .293/-.195/-1.37 		& $8.6$	\\
%%			& $\eta'$	& .595/1.85/-.533 		& -7.2	\\
%\hline
%\end{tabular}
%\end{table}

\begin{table}
\caption{
Summary of the couplings of pseudo-scalar color-singlet state $a$ in the considered composite models. $c_t$ and $c_g$ are given in units of $v/F_\pi$. $c_t$ is shown for the three benchmarks $(n_\psi,n_\xi)=(2,0)/(-4,2)/(4,2)$. }
\label{tab:models}
\begin{tabular}{ l c c }
 model 		& $\quad c_t[v/F_\pi]\quad$ & $c_g[v/F_\pi]$ \\
 \hline\hline
M3			& 0.934/1.09/-2.65		& 5.44	\\
M8			& 0.926 /1.54/-2.16		& 1.54	\\
M9		    & 0.293/-0.195/-1.37 		& 8.6	\\
\hline
\end{tabular}
\end{table}

In Fig.~\ref{fig:Exclm1500-model} we show the value of $c_t$ and $c_g$ for each model together with the exclusion region (above the black curve) for a fixed mass $ m_a=1.5\,{\rm \TeV}$. We consider an integrated luminosity of $L=20\ifb$ for the exclusion limit and a systematic error $\epsilon_{\rm SYS}=5\%$.
The different line colors in the figure refer to the different models: red is M8, yellow M9 and brown M3. The styles of the lines represent the fermionic charges: $(n_\psi,n_\xi)=(2,0)$ (solid line), (-4,2) (dashed) and (4,2) (dot-dashed). 
%Finally, for each of this scenario we show the couplings of $a$ state for the case $\alpha=\zeta$ (cyan dots) and of the $\eta'$ state for the case $\alpha=\zeta/2$ (magenta dots) 
Each line scans the values of $F_\pi$ from $v$ (most external and largest couplings) to $8v$ (most internal and smallest couplings), the dots represent the values     $F_\pi=n\,v$, with $n$ an integer between 1 and 8 included.
Also shown for reference in the upper region the couplings of the $\eta_{63}$ state of the Fahri-Susskind one-family model~\cite{Farhi:1980xs}.

From the figure we can get the minimal value of the compositeness scale $F_\pi>F_\pi^{min}$ for which state $a$ would still not have been observed for different scenarios. For instance, for model M8 (red lines), $v\lesssim F_\pi^{min}\lesssim 2v$ depending on the values of $(n_\psi,n_\xi)$. The model M3 is more constrained, and $6v\lesssim F_\pi^{min}\lesssim 7v$ for (-4,2) and $F_\pi^{min}\sim 5v$ for (2,0) or (4,2).
Model M9 has low values of $c_t$ but values $ F_\pi\gtrsim 6v$ can be excluded  for the case  (-4,2), while the other scenarios are hard to access in the $\ttb$ search.

Other decay channels have been analyzed in Ref.~\cite{Belyaev:2016ftv}.

\begin{figure}[H]
\begin{center}
\includegraphics[width=0.8\textwidth]{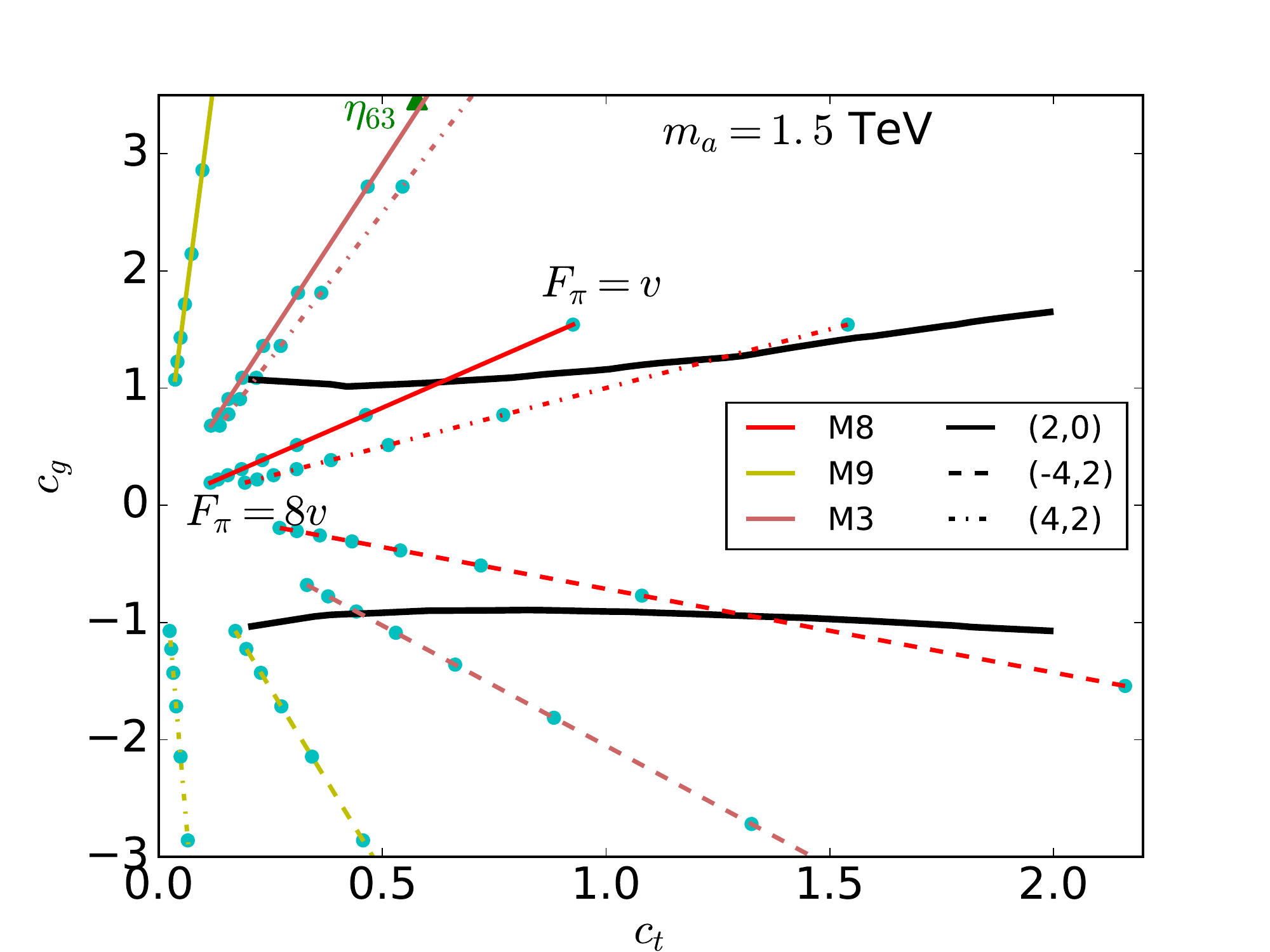}  
\end{center}
\caption{In black thick lines the exclusion limits for the partial-compositeness models considered here are drawn. An integrated luminosity of $L=20\ifb$ and an
  uncertainty of $\epsilon_{\rm SYS}=5\%$ are assumed. The model lines refer to models M8 (in red), M9 (yellow) and M3 (brown)
  introduced in the text. The styles of the lines represent the fermionic charges: $(n_\psi,n_\xi)=(2,0)$ (solid line),
  (-4,2) (dashed) and (4,2) (dot-dashed). Each line scans the values of $F_\pi$ from $v$ (most external and largest couplings)
  to $8v$ (most internal and smallest couplings), the dots represent the values $F_\pi=n\,v$, with $n$ an integer between 1 and 8 included.}
\label{fig:Exclm1500-model}
\end{figure}

%%%%%%%%%%%%%%%%%%
\subsection{Broad scalar color singlet}

In this benchmark scenario we assume a CP-even color-singlet scalar that can, apart from top quarks and gluons, also decay into other particles and
is thus much broader than the previous scenarios. We choose a total width of 20\% of the resonance mass $\Gamma_\sigma=20\%\, m_\sigma$.
The rationale for choosing a larger width is the fact that the scalar tends to decay also to weak bosons. Indeed, we expect a large sensitivity in this decay channel which might be competitive w.r.t. top pair production.

In this scenario the signal is very weak and thus hard to be observed unless the systematic uncertainty is improved to values below
5\% or higher values of $c_g>3$ are considered.  In Fig.~\ref{fig:scalar} on the left we show the line-shape for $m_\sigma=900\,{\rm \GeV}$, $c_t=c_g=1$. It can be noticed that the yields are always below 5\%. On the right panel we show the $\chi^2_{10}=2$ contours in the $(m_\eta,c_g)$ parameter space plane  for $c_t=1$. Varying the assumed systematic uncertainties between $\epsilon_{\rm SYS}=1\%-2\%$ determines the band of the exclusion limit.
The integrated luminosities are $L=20\ifb$ (blue line) and $L=300\ifb$ (black). Limits are given considering interference (dashed lines)
and neglecting it (solid lines). A large interference effect can be noticed, which is in fact larger than the pure signal.

\begin{figure}[H]
\includegraphics[width=0.49\textwidth]{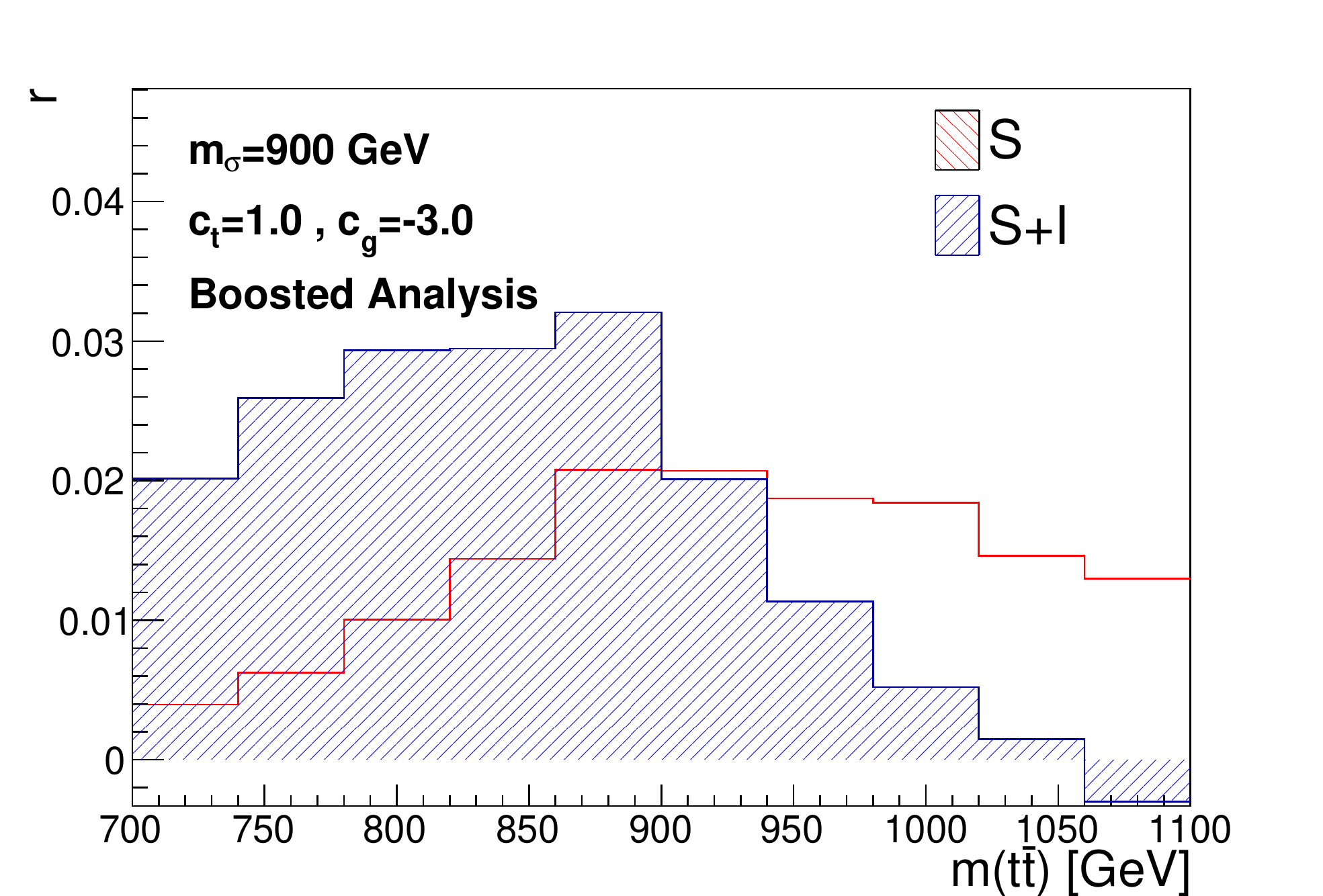}  
\includegraphics[width=0.49\textwidth]{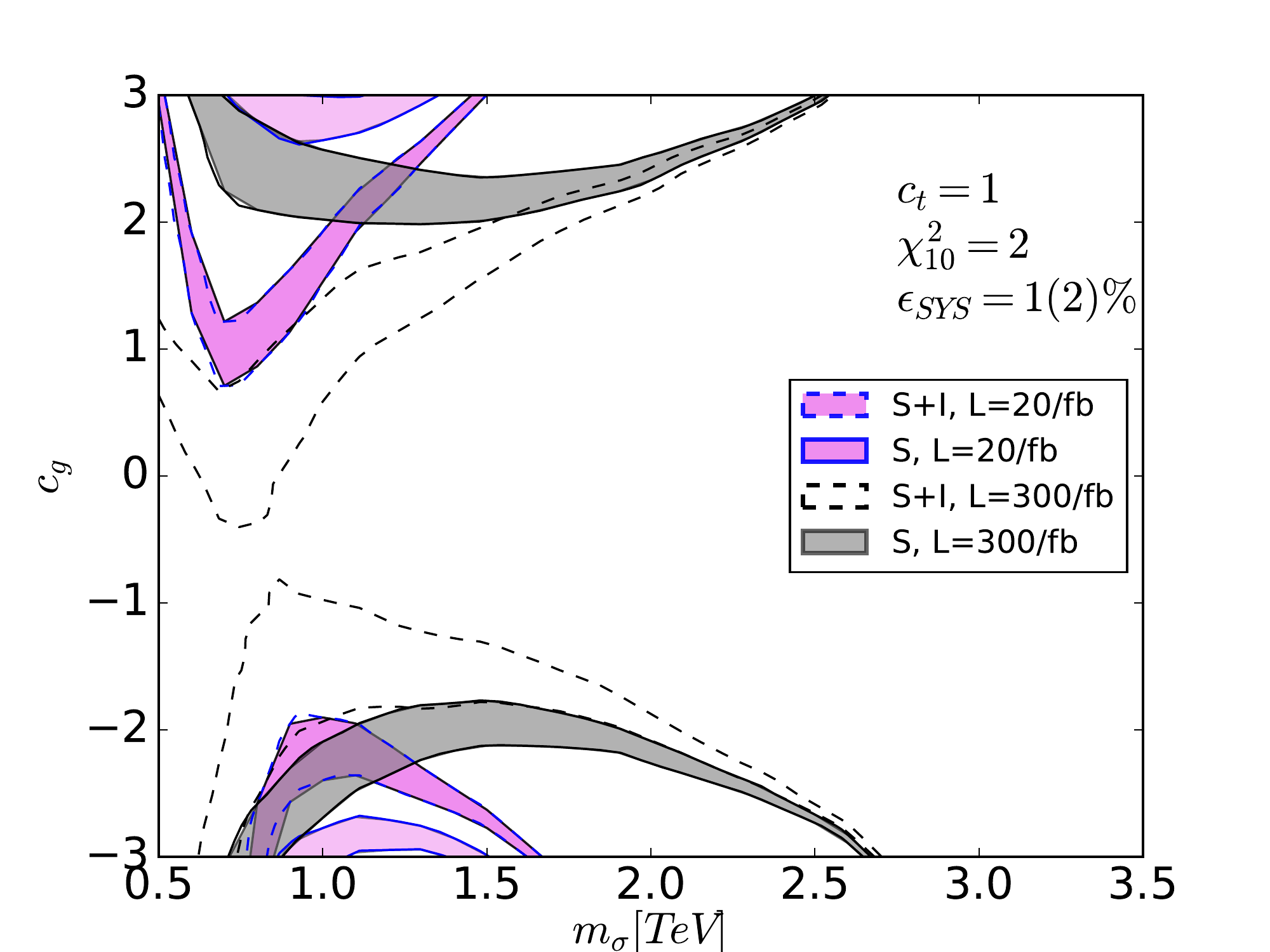}  
\caption{\emph{Left:} Normalized top-pair mass distributions $r$ for a color-singlet scalar with $c_g=c_t=1$, $m_\sigma=900\,{\rm \GeV}$
  and $\Gamma_\sigma=20\%\,m_\sigma$.
  \emph{Right:} Exclusion limit ($\chi^2=2$) in the $(m_\sigma,c_g)$ parameter space and $c_t=1$ for such scalar state. The color scheme is the same as in Fig.~\ref{fig:Exclct1}. In both panels the \emph{Boosted} analysis have been adopted. }
\label{fig:scalar}
\end{figure}

\section{Conclusion}
\label{sec:conclusion}

In this work we have provided a framework to reinterpret the SM $\ttb$ differential cross section
measurements in terms of exclusion limits for signatures of NP scalar resonances decaying into
$\ttb$. The method relies on the detailed simulation of the SM prediction at particle level
with the \Sherpa\ Monte Carlo, the subsequent analysis in the \Rivet\ framework, which can be
directly compared with the measured distributions provided by the experimental collaborations,
a modeling of the NP scenarios efficient enough to allow a scan over a large range in parameter
space, and finally a statistical analysis to determine the excluded regions.

In the simulation of top-pair production we take into account higher-order QCD corrections through
matching LO or NLO matrix elements to parton showers and merging partonic processes of varying
multiplicity. To validate our simulation we compare to data from the ATLAS collaboration, finding
very good agreement. As New Physics contributions we consider CP-even and CP-odd scalar resonances,
being either color-singlets or octets. To model the signal we devise an efficient and fast
\emph{reweighting} method allowing to scan large regions of parameter space without the need of
full re-simulation and re-analysis for each parameter point. For our simplified model we have
derived exclusion limits based on a simple $\chi^2$ analysis, that can subsequently be used to
set limits on other specific models, and we consider a model of partial compositness as an example. 
We showed the importance of properly accounting for interference between the New Physics signal and
the SM background in setting the exclusion limit, as well as of using a full line-shape analysis
which is not necessarily a simple Breit-Wigner shape due to the interference effects.

By confronting SM precision measurements with hypotheses for New Physics models stringent exclusion
limits on the parameters of the latter can be obtained, providing complementary sensitivity to
direct searches. The methodology laid out here can be readily applied to other observables than the
top-pair invariant mass considered here. It relies on a solid understanding of the respective
SM expectation and the uncertainties related to the theoretical predictions and the experimental
data.   

\section*{Acknowledgments}

This work was supported by the European Union through the FP7 training network MCnetITN
(PITN-GA-2012-315877) and the Horizon2020 Marie Sk{\l}odowska-Curie network MCnetITN3 (722104).
Federica Fabbri especially wants to thank MCnetITN for the opportunity to hold a short-term
studentship at the II. Institute for Physics at G\"ottingen University. 

\appendix
    \section{Line-shapes samples}
\label{app:lineshapes}

\begin{figure}[H]
\includegraphics[width=0.49\textwidth]{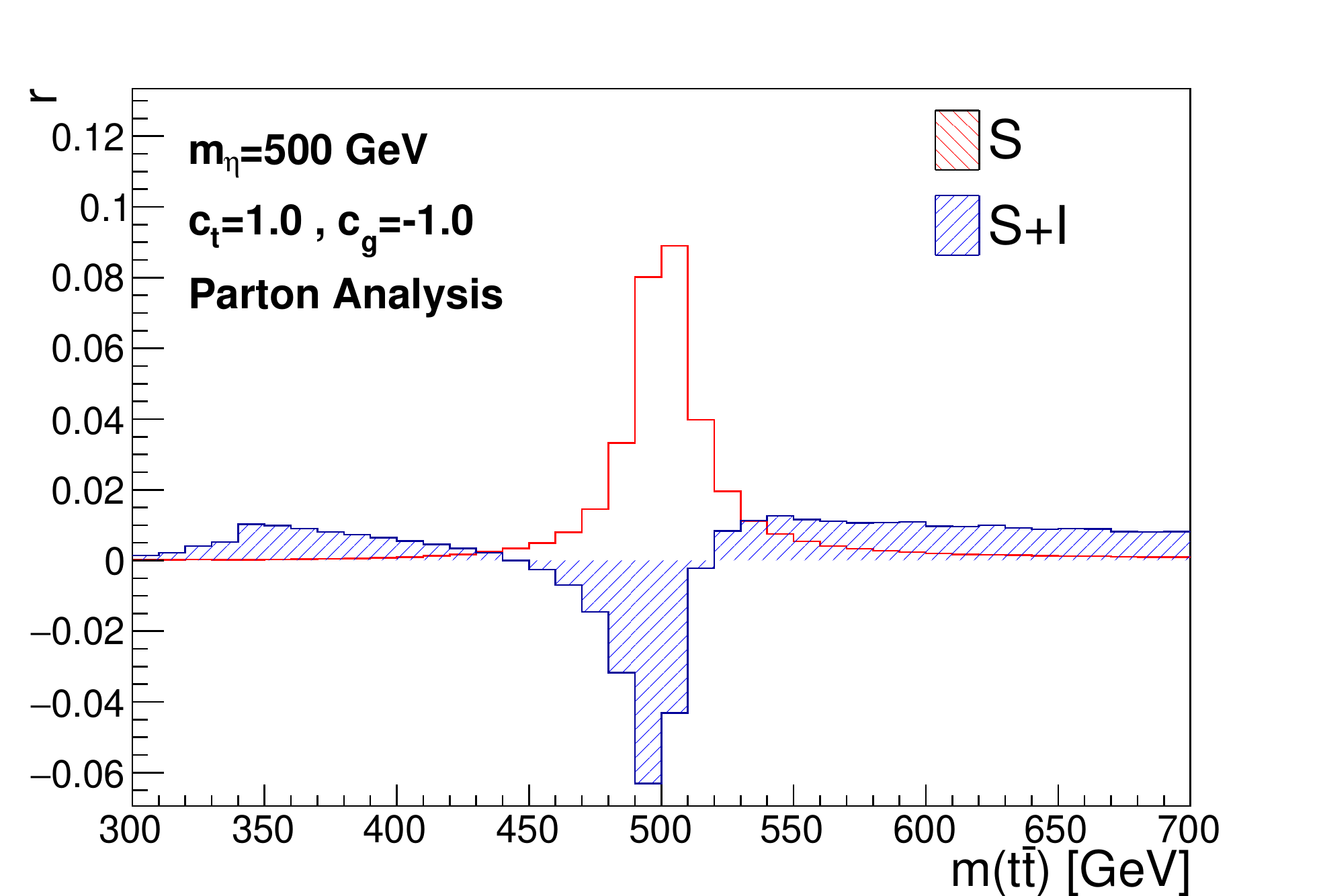}
\includegraphics[width=0.49\textwidth]{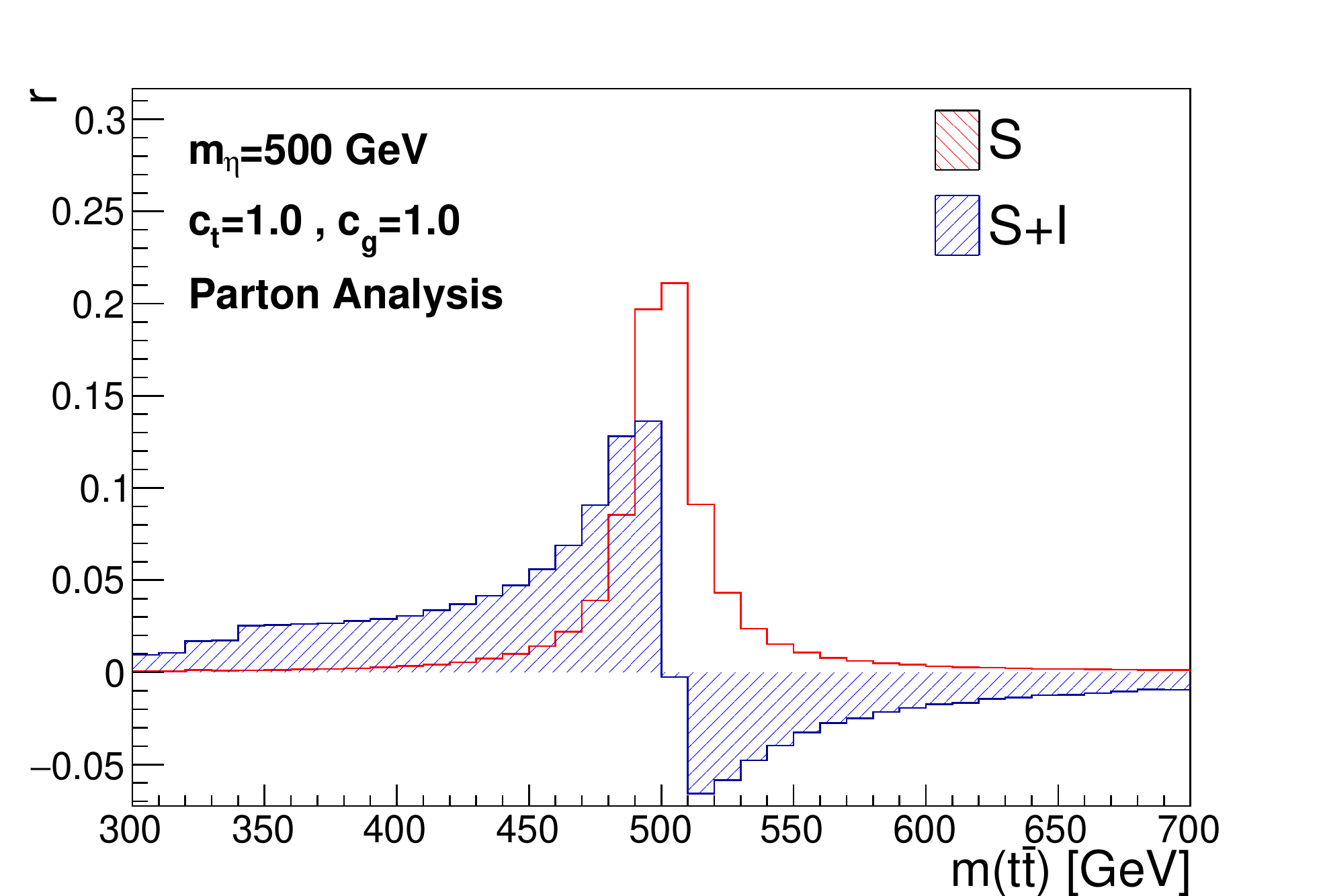} 
\caption{Normalized top-pair mass distributions, $r_H=\frac{d\sigma(H)/dm}{d\sigma_{\rm SM}/dm}$ ($H=S,S+I$), for
  a pseudo-scalar signal ($m_\eta=500$ GeV) in the \emph{Parton Analysis}.}
\end{figure}

\begin{figure}[H]
\includegraphics[width=0.49\textwidth]{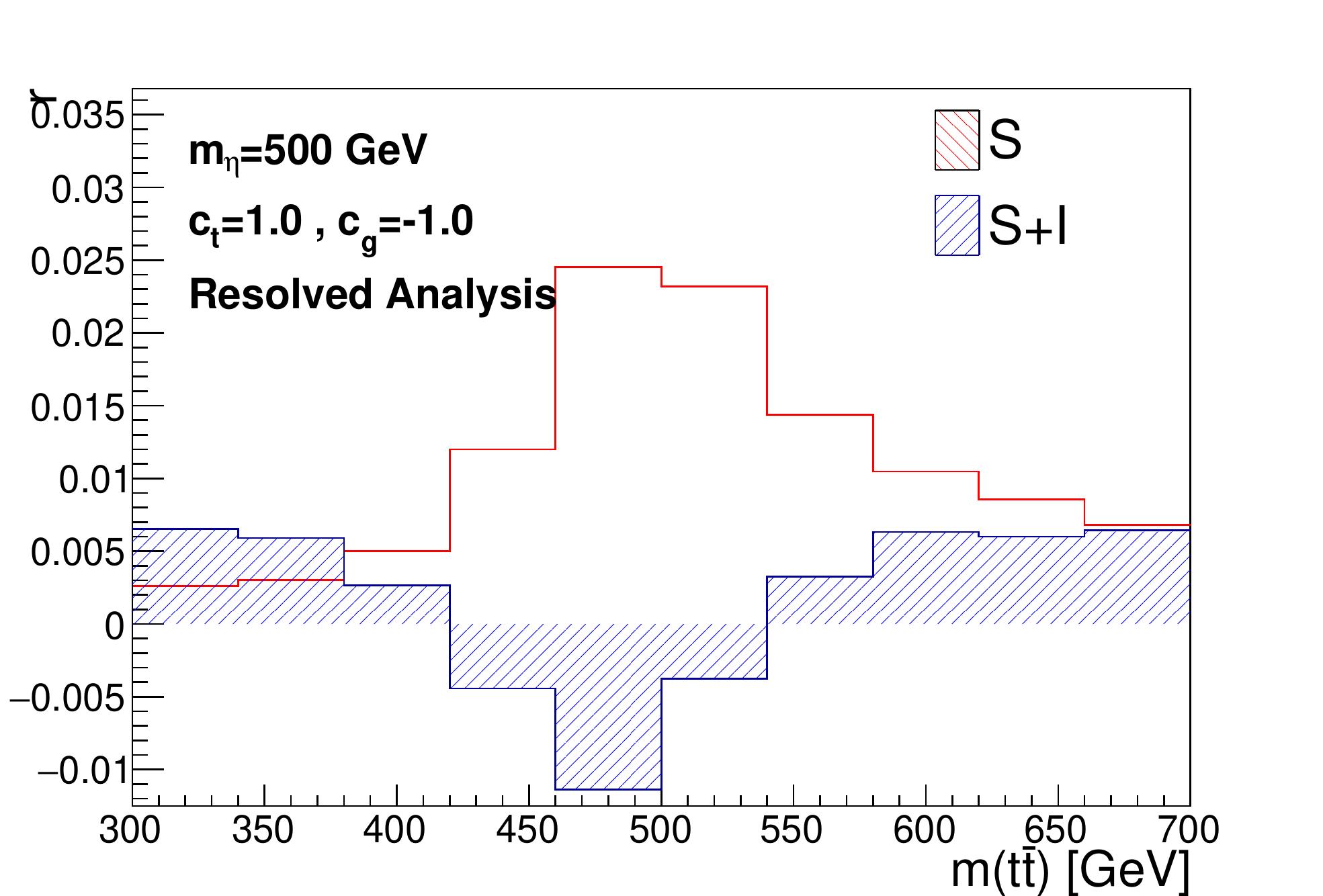}
\includegraphics[width=0.49\textwidth]{{plots_ScanPct1_13TeV_d10e1noMCL20_cResolved129_m500ct1.0cg1.0}.pdf} 
\caption{Normalized top-pair mass distributions, $r_H=\frac{d\sigma(H)/dm}{d\sigma_{\rm SM}/dm}$ ($H=S,S+I$), for
  a pseudo-scalar signal ($m_\eta=500$ GeV) in the \emph{Resolved Analysis}.}
\end{figure}

\bibliographystyle{JHEP}

\bibliography{ttbar-interference}

\end{document}
\grid